\documentstyle[epsfig]{cupconf}

\ifoldfss
\else
  \ifnfssone
    \newmathalphabet{\mathit}
      \addtoversion{normal}{\mathit}{cmr}{m}{it}
      \addtoversion{bold}{\mathit}{cmr}{bx}{it}
    \newmathalphabet{\mathcal}
      \addtoversion{normal}{\mathcal}{cmsy}{m}{n}
    \else
    \ifnfsstwo
    \fi
  \fi
\fi

\def\gtwid{\mathrel{\raise.3ex\hbox{$>$\kern-.75em\lower1ex\hbox{$\sim$}}}}
\def\ltwid{\mathrel{\raise.3ex\hbox{$<$\kern-.75em\lower1ex\hbox{$\sim$}}}}

\def\hexnumber#1{\ifcase#1 0\or1\or2\or3\or4\or5\or6\or7\or8\or9\or
 A\or B\or C\or D\or E\or F\fi }

\makeatletter
\ifx\CUP@mtlplain@loaded\undefined
\else
\fi
\makeatother
 \makeatletter
 \ifx\CUP@mtlplain@loaded\undefined
   \font\tenbmi=cmmib10 at 10pt
   \font\sevenbmi=cmmib10 at 7pt
   \font\fivebmi=cmmib10 at 5pt

   \newfam\bmifam
   \textfont\bmifam=\tenbmi
   \scriptfont\bmifam=\sevenbmi
   \scriptscriptfont\bmifam=\fivebmi
   
 \fi
 \makeatother
\ifnfsstwo

\fi
\ifnfssone

\fi
\ifoldfss

\fi

\mathchardef\varLambda="0103

\makeatletter
\ifx\CUP@mtlplain@loaded\undefined
\else
\fi
\makeatother
\makeatletter
\ifx\CUP@mtlplain@loaded\undefined
  \font\tenbms=cmbsy10
  \font\sevenbms=cmbsy10 at 7pt
  \font\fivebms=cmbsy10 at 5pt
  \newfam\bmsfam
  \textfont\bmsfam=\tenbms
  \scriptfont\bmsfam=\sevenbms
  \scriptscriptfont\bmsfam=\fivebms

  \edef\bsy@{\hexnumber\bmsfam}
  \mathchardef\bnabla="0\bsy@72
\fi
\makeatother
\title[Polarization Insights for AGN]{Polarization Insights for Active
Galactic Nuclei}

\author[R. R. J. Antonucci]%
{R\ls O\ls B\ls E\ls R\ls T\ns
 A\ls N\ls T\ls O\ls N\ls U\ls C\ls C\ls I$^1$\ns}

\affiliation{$^1$Department of Physics, University of California,
Santa Barbara, CA 93106-9530, USA}

\begin{document}
\ifnfssone
\else
  \ifnfsstwo
  \else
    \ifoldfss
      \let\mathcal\cal
      \let\mathrm\rm
      \let\mathsf\sf
    \fi
  \fi
\fi

\maketitle

\begin{abstract}
Optical spectropolarimetry and broadband polarimetry in other
wavebands has been a key to understanding many diverse aspects
of AGN.  In some cases polarization is due to synchrotron radiation, 
and in other cases it's due to scattering.  Recognition of relativistically
beamed optical synchrotron emission by polarization was vital
for understanding blazars (BL Lacs and Optically Violently
Variable quasars), both physically and geometrically.  Radio
polarimetry of quiescent AGN is equally important, again for both
purposes.  Scattering polarization was central to the Unified
Model for Seyferts, Radio Galaxies and (high ionization) Ultraluminous
Infrared Galaxies.  It provides a periscope for viewing AGN from other
directions.  Finally, if we could understand its message, polarization
would also provide major insights regarding the nature of the AGN
``Featureless Continuum" and Broad (emission) Line Region.

I point out that high ionization ULIRGs have all the exact right
properties to be called Quasar 2s.  Mid-IR observations generally
don't penetrate to the nucleus, greatly reducing their ability
to diagnose the energy source.  In particular, LINER ULIRGs
aren't necessarily starburst-dominated, as has been claimed.

\end{abstract}

\firstsection % if your document starts with a section,
              % remove some space above using this command.
\section{Seyfert Galaxies}\label{sect:1}

\subsection{Type 2 Seyferts}

\subsubsection{Polarization alignments and hidden Type 1 Seyfert nuclei}

In the 1970s the continua of Seyfert 2s were decomposed 
into two parts:  relatively red light from the old stellar population,
and a bluer component modeled satisfactorily with a power law.  The
latter was called the ``Featureless Continuum," in a commendable
attempt to avoid prejudice as to its nature.  (Unfortunately some of
them were later found to have strong features;  see below.)

The small ($\sim1$\%) V-band polarization often seen usually derives from
the power law component.  Most of these continua are dominated by unpolarized
starlight, so the implied ``FC" polarization is sometimes intrinsically large.
The red starlight also strongly affects the wavelength-dependence of P so it really
had to be removed.  It was shown that for the brightest and best observed
Seyfert 2, NGC1068, the true FC polarization was a surprisingly high 16\%,
and independent of wavelength (\cite{Miller83}; \cite{McLean83}).
The former paper noted the cause might be either scattering above and below
an opaque torus, or synchrotron emission;  the latter focused on synchrotron
radiation as the more likely.

Following the initial discovery of a geometrical relationship 
between optical polarization and radio axis for quasars (\cite{Stockman79}),
\cite{Martin83} and I (Antonucci 1982a, 1983) sought
such patterns in Seyferts and radio galaxies.  (Ulvestad and Wilson 
were discovering tiny, weak, but linear radio sources in many Seyferts, 
using the new Very Large Array.)  The Martin {\it et al.\/} paper
presented a lot of data, but didn't find alignment effects.  I did
claim to see them, and with essentially the same data.  I think there were
two reasons for the difference: 1) I didn't consider the whole sample
statistically, but
only the few whose polarization was very likely to be intrinsic to
the nuclei, and 2) I divided them into the two spectroscopic classes
(Type 1 and 2), considering each separately. 

There was pretty good evidence that the Type 2s tended to be polarized
perpendicular to the radio axis, and not so good evidence that the
Type 1s were parallel.  (I tended to believe the latter though, because
of the parallel polarizations of their ``cousins"(?), the radio loud
quasars.)

\begin{figure} 
  \centerline{\epsfig{file=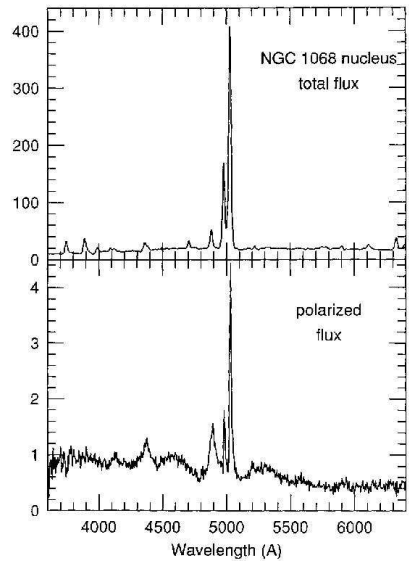,height=3.0in}}
  \caption{Total and Polarized Flux Spectra of NGC 1068.} 
  \label{fig:1}
\end{figure}

At the same time Joe Miller and I (I was the thesis student) were
observing NGC1068 and the radio galaxy 3C234 with a new spectropolarimeter.
This device had just been built, mainly by J Miller and G Schmidt.  (Since 
I have no talent with instrumentation, my goal was just to avoid
breaking it.)  I realized pretty quickly that the explanation for
the high ``perpendicular" polarization of 
3C234 was reflection from a quasar hidden inside a torus (Antonucci
1982a, b).  We had to puzzle over NGC1068 longer, I suppose
because the strong starlight confused us.  Because of the wavelength
independence of starlight-subtracted P, as well as some other moderately
good reasons, we thought the scattering was by free electrons.  Miller
{\it et al.\/} (1991) then
presented much better data, and showed the polarized
spectrum {\it as scattered off dust clouds in the host galaxy}\/.
%AS SCATTERED OFF DUST CLOUDS IN THE HOST GALAXY.
These data indicated that 1) the dust-scattered light is much bluer than
the inner, putative electron scattered light;  and 2)  the
broad lines are narrower, indicating that the line widths
are somewhat smaller than those in the electron-scattered polarized flux plot,
when the reflection is from dust.  This
confirms that the nuclear scatters are electrons, and provides
a temperature estimate of $\sim300,000$ K.  Note that according to the
true line widths, NGC1068 would be classified as a ``Narrow Line
Seyfert 1" by astronomers looking from above.  The Miller {\it et al.}\/ 1991
nuclear data appear as the present Fig.~\ref{fig:1}.

We now know that many Seyfert 2s are just Seyfert 1s hidden inside
opaque tori, and that therefore at least part of the class difference
is just orientation.  Some early references are Miller and Goodrich 1990 and
Tran {\it et al.\/} 1992.  We still don't know if this applies to all the 2s.
For a while I thought it probably didn't because Kay (1994) was
finding that the FC for many Seyfert 2s have low polarization.  However,
starting with 3C234, it was becoming clear that some of that blue
light was coming from hot stars or some other source, and that
there was in fact highly polarized reflected light from hidden 1s
in many more (Miller and Goodrich et al 1990, Tran 1995a,b).
The high polarization of the 
actual scattered part of the FC is proven by the polarization
of the broad lines, and the normal Seyfert 1 spectra in polarized flux.
(The key here is that the broad lines have normal equivalent width
in polarized flux.)
In most cases we just have very high $\sim10$--20\% lower limits on the
broad line region (BLR)
polarization since at least I can't see them at all in the total flux in
many objects (P = polarized flux in broad line/total flux in broad
line.)  Soon afterwards Heckman {\it et al.\/} 1995,
and Gonzalez-Delgado {\it et al.\/} 1998
showed definitive spectroscopic evidence that most of the FC in several
famous objects is in fact light from hot stars.

As noted, the general applicability of the hidden-1 model for the 2s
is still poorly known.  It would be great to select a sample by some
nearly isotropic property such as $60\mu$ emission, and observe them
all down to a certain level of sensitivity.  One (contentious) idea
is that the scattering regions are themselves often partially occulted, so that
the trick works only for those viewed at relatively small inclinations
(Miller and Goodrich 1990).  Models of the torus (e.g., Pier and
Krolik 1993) indicate that such relatively polar views would expose
warm dust to the observer, and it's been argued that we can in fact
detect the hidden 1s in all the {\it warm}\/ Seyfert 2s (Heisler
{\it et al.\/} 1997).

\subsubsection{Mirror and torus;  a 3-D image of an AGN}

These components were entirely hypothetical in the early 1980s.
The mirrors (scattering regions) weren't resolved significantly
from the ground (but see Elvius 1978 for resolution of the outer
dust-scattering part of the NGC1068 mirror).  HST however
could spatially resolve some of them.  Our multiaperture
HST UV spectropolarimetry resolved the inner ``electron-scattering"
mirror in NGC1068 (Antonucci {\it et al.\/} 1994),
as did the beautiful polarization images by Capetti {\it et al.\/} (1995a,b).

The central arcsec or so shows neutral (wavelength-independent)
scattering, and for this and other reasons, 
electron scattering seems to dominate there.
The $\sim400$km/sec redshift of the broad lines in polarized light
indicates polar outflow; recall that the scattering must be polar to
explain the position angles (PAs).  The dominance of electron scattering
means that this gas has lost virtually {\it all}\/ its dust, probably by
travelling inside the sublimation radius ($\ltwid1$pc).  Finally, as noted
above, the gas temperature is thought to be $\sim300,000$K because 
the electron-scattered
versions of the broad lines are somewhat wider than the lines
seen scattered off dust clouds in the host galaxy (Miller, Goodrich 
and Mathews 1991).

The probable physical basis of this whole
occultation/reflection scenario was first provided by
Krolik and Begelman (1986, 1988).  Krolik and collaborators
also calculated the theoretical
requirements and consequences of the scenario, predicting,
for example, that the scattering region should produce a
high-ionization Fe K-$\alpha$ line of enormous ($>1$keV) equivalent
width, as observed.  The large $\sim1${\tt "} $=75$pc
size of the electron scattering mirror
was anticipated by the models of Miller, Goodrich and
Mathews (1991).

\begin{figure} 
  \centerline{\epsfig{file=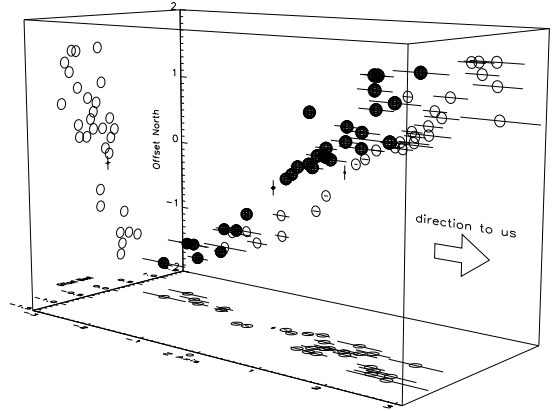,height=2.5in}}
  \caption{Three dimensional inversion of the electron-scattering region in
NGC1068.} 
  \label{fig:fg12}
\end{figure}

Capetti {\it et al.\/} did a fine job analyzing the HST imaging,
delineating for example the inner electron-scattering
(neutral scattering) mirror, and the outer regions which
show dust scattering (strong rise in cross-section with
frequency).  There was a slight puzzle left over from
their analysis:  in the inner region the polarization
PAs were not quite centrosymmetric as expected
for scattering.  In principle this means that the
hidden source isn't quite pointlike.  However, Kishimoto (1999 and p.c., 2000)
found that the deviations were entirely traceable to
instrumental effects.  Also, since we know the polarization
phase function perfectly for electron scattering, it's
possible to determine the angle between the nucleus and
a scattering cloud, relative to the sky plane.  For example,
if some scattered light has a polarization of 100\%, we
know it's right-angle scattering, so that the cloud is
right in the sky plane.  Kishimoto's paper shows the first
(I think) 3-D image of an AGN (Fig.~\ref{fig:fg12})!  See that paper 
for some caveats.  The overall image is fairly robust 
I think, and wonderful.

I won't say much about the earliest detections of the torus material since
they didn't involve polarimetry.  But structures which
are acceptable manifestations of it have been observed
in the infrared, and in various molecular emission lines.  
The observed molecular
tori, whose ``outer radii'' are $\sim100$pc, are observed in NGC1068,
and also in luminous quasars and infrared galaxies (Section~\ref{sect:3}
has some more information on this).  Two
points to bear in mind:  the polarization and spectral-energy
distribution information indicate that the {\it inner}\/ radius
is somewhere near the sublimation radius.  Independent
of the unified model, the near-IR upturn in the SEDs alone
implies a substantial covering factor of hot dust.  And
remember that the polarimetry indicates a torus only in the
sense of something opaque, with holes along the axis
so the photons can get out.  In the real world there could
be (and are) more complicated structures, including bars;
it's also possible that a warped thin disk could do the job,
but it would have to be extremely warped and ``tall" to 
simulate the obscuring behavior of a torus.

\subsection{Type 1 Seyferts}

\subsubsection{Intrinsic Nuclear Polarization}

Type 1 Seyfert nuclei generally have low intrinsic polarization ($\ltwid2$\%),
so it is often overwhelmed by dichroic or other processes
on the $\sim$kpc scale.  This is apparent when the narrow line region (NLR)
and continuum polarizations agree, and/or both have PAs parallel to the host
major axis in a high-inclination galaxy
(Thompson and Martin 1988).  If in doubt, a polarization image
can help determine whether or not the 
polarization is associated with the nucleus.

There is fairly good evidence that the nuclear polarization PAs
tend to align with the axes of the nuclear radio sources.  As with
the 2s, considerable 
uncertainty comes from the curved nature of the latter.
Many years ago the available small sample seemed to show the alignment
(Antonucci 1983).  Since then various people have updated it, and 
I haven't always agreed with what was done.  Here is my personal
update:

The original criteria were: 

\begin{enumerate}
 \item $P\gtwid0.6$\%
 \item sky position in a {\it good}\/ Galactic interstellar polarization zone.

This means it lies 
within a personal
pre-defined sky area with relatively little expected interstellar
polarization.  I'd chosen these areas using the maps in
Mathewson and Ford 1970.
 \item linear radio source.  
\end{enumerate}

\vskip 2.5mm
Updates relative to the 1983 paper:

\begin{enumerate}
 \item I'd drop 3C390.3 on the grounds that it's a radio galaxy;  I was young
and dumb in 1983 so I was
following the classifications in an earlier paper by someone else. 

 \item Drop NGC3227 because the [O III] data show that the polarization
arises in the host galaxy rather than in the nucleus\
(Thompson {\it et al.\/} 1980).

 \item Add Mrk509:  Polarization PA: $\sim1.0$\%, at 130--150
(true variability) degrees: see references 
                 in Singh and Westergaard 1992.  The 
                 radio PA is $130\pm10$  (Singh and Westergaard 1992):
                 nearly parallel

 \item Mrk704:  Delta PA $= 81^\circ$!  from Martel 1996;  also in Goodrich
    and Miller 1994  

 \item Mrk1048  Delta PA $= 12^\circ$!  from Martel 1996;  also in Goodrich
    and Miller 1994

 \item NGC3516  Polarization $\sim0.8$\% at $\sim2^\circ$;  
             Radio 0--10;  See Miyaji {\it et al.}\/ 1992: parallel.

 \item NGC5548  Delta PA = $\sim55$:  Martel 1996

 \item Mrk9        Delta PA = $\sim28$:  Martel 1996

 \item Mrk304      Delta PA = $\sim90!$:  Martel 1996: perpendicular.

% \item Mrk766   $P = 2.34\pm0.02$ at $90\pm0.3$  (\cite{Goodrich89};
%see also Ulvestad {\it et al.\/} 1995).

% \item Mrk1126 = NGC7450  $P = 0.47\pm0.04$ at $173\pm3$  (\cite{Goodrich89}; 
%        see also Ulvestad {\it et al.\/} 1995). Radio PA $\sim100$ but not
%        considered very robust.

 \item Mrk957     $P = 0.62\pm0.06$  at $43\pm3$  (Goodrich 1989)
               Radio PA $50\pm10$;
               see also Ulvestad {\it et al.\/} 1995)  
\end{enumerate}

While the data are still marginal, an optimistic and plausible
summary is this.  Most Seyfert 1 nuclei show optical polarization
parallel to the radio axis;  the exceptions may favor a perpendicular
relationship.

\subsubsection{Variety of BLR behavior.  Constraints on the nature of the
underlying continuum}

In all known cases, the BLR line polarization is less than or
about equal to that of the continuum, in magnitude.  The action
within the line profiles depends on the object.  Often both the
magnitude and the angle vary rapidly across the profile. It is
undoubtedly encoded with lots of information on the nature of
the BLR, but it's very difficult to decode.  The most heroic 
and intriguing attempt is Martel's (1998) analysis of the broad
H-alpha line in NGC4151.  

There are several c.~1980 papers by the Steward group (R Angel,
I Thompson, E Beaver, H Stockman, and probably some others).
These authors were the pioneers measuring Seyfert polarization.
%but the papers are before the ``modern era"!
Later (CCD) data supercede their observations.

NGC4151 was observed by Schmidt and Miller (1980), who found that
the integrated broad line polarization was undetectably low with their
data quality.  There are some quasars that also have undetectably
polarized broad lines.  This is of great interest because it
means that the polarized flux plot looks like a noisy version
of the spectrum, but with {\it the broad lines and small blue
bump (Fe II plus Ba continuum from the BLR) scraped off
as by a razor!}\/  See e.g., Antonucci 1988 where I present some data
``borrowed" from Miller and Goodrich (pc); and Schmidt and Smith 2000.
There are a few others.

Thus we can tell what the underlying 
continuum is doing\dots at for example, the Ba edge.  The answer is:
nothing.  (Nothing happens at the Ly edge either.)  This is
not trivial.  Models such as accretion disks which are cool
enough to match the optical slopes would at least na\"\i vely
show the Ba edge in absorption.  In one very important regard
the Ba edge is more interpretable than the Ly edge:  the relativisitic
smoothing effects at the Ly edge are much reduced.  If there's a feature,
we should see it!

\begin{figure} 
\centerline{\epsfig{file=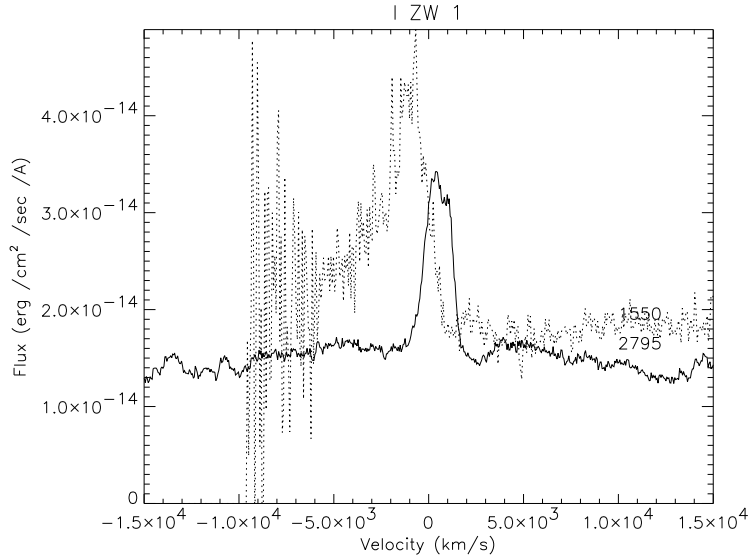,height=3in}}
\caption{Overplot of C IV and Mg II broad line profiles in the Narrow Line
Seyfert 1 or Quasar called I Zw 1, M Kishimoto 2000 p.c.  This figure strongly
suggests that C IV (and other high-ionization lines) derive from a wind,
whereas Mg II (and the other low-ionization lines) do not.  For a detailed
analysis of I Zw 1, see Laor {\it et al.}\/ 1997.}
  \label{fig:plot}
\end{figure}

We were asked to suggest observing projects for young astronomers,
and I think the following is quite good if practical.
The high-ionization lines in ``Narrow Line''
Seyfert 1s are often blueshifted and very broad relative to the
low ionization lines and the systemic velocities.  (As far as I know
this important result was first shown by
Rodriguez-Pascual {\it et al.\/} (1997),
though you wouldn't get that impression from some more recent papers!)
To tempt the reader, I show the spectacular example of I Zw 1
in Fig.~\ref{fig:plot}.
(Some of those in Leighly 2000 are even better.)
This was given to me by M.~Kishimoto.  I needed no convincing after 
seeing that plot, that the high-ionization lines are from a wind,
in that object and maybe others.  Early proponents such as \cite{Collin88}
deserve congratulations.  It would be great to measure
the polarization behavior of the two types of line.  There are at present no
space spectropolarimeters, so what's needed is a bright object with 
enough redshift to bring a high ionization line like C IV 1549 into
the optical --- can we find such an object that shows the difference
in the line profiles clearly?

Recovering from that digression, I just refer you to these
papers on Seyfert 1 polarization:
Berriman (1989) interpreted
his broadband Seyfert 1 survey data as indicating dust scattering
for most objects.  More data including spectropolarimetry were presented
by Brindle (1990).  J.~Miller and students obtained some fairly
large (thesis-size) data sets (Goodrich and Miller 1994;
Martel 1996).  These papers show in detail the complex
behavior inside the broad emission lines.

\subsection{Broad Absorption Line Quasars}

This subject has recently been reviewed by Schmidt and Hines 1999,
and by Ogle et al.\ 1999, so here I'll just present the main points.
See also Hutsemekers and Lamy 2000 for more modern-quality data.

1.  BAL quasars comprise $\sim10$\% of optically selected quasars.  
The UV absorption from the troughs and possible accompanying dust
absorption make them hard to find optically, so their true
incidence is thought to be $\sim30$\%.   For two takes on this, see Goodrich
1997 and Krolik and Voit 1998.  Goodrich points out that the significant
electron scattering optical depth inferred for some BALs would
itself cause the continuum to be emitted anisotropically.

2.  These are really ``broad scattering quasars":  The troughs
really represent resonant scattering of heavy element lines in the UV.
With unit covering factor (and no dust, etc!) the photons should all
be returned to the beam and appear as emission line flux.  However
the observed emission lines are much too weak for that to be correct.  In fact,
the emission line strengths are fairly normal, so much of that
emission must be from the usual collisional excitation in the broad emission
line region.  Thus the BAL covering factor is thought to be
much less than unity.

3.  Points 1 and 2 require that there are many objects intrinsically
the same as BAL quasars which do not manifest as such from our sight
line.  With this semi-quantitative reasoning, all radio quiet
quasars may be intrinsically very similar, appearing as BALs
from some directions.

4.  Because non-spherically symmetric resonant scattering is
involved, the emission lines were predicted to be polarized
(Scargle et al.\ 1970), but this expectation hasn't generally been realized
(Stockman et al.\ 1981 for the earliest data;  Ogle et al.\ 1999 for the
current state of the art).
There are in several cases apparent manifestations of
some resonant-scattered emission line photons in the polarized flux spectra.
But it's safe to say the polarization behavior of the emission lines
is poorly understood.

5.  Among radio loud objects, troughs of somewhat lesser widths
are sometimes seen (e.g., Antonucci et al.\ 1993 on OI 287).  Recently the
``FIRST" radio survey has turned up some ``radio-intermediate"
BAL quasars, and at least one classical double (Gregg et al.\ 2000)!
I think the incidence of truly broad troughs is still much lower
among the powerful radio sources.  

  Members of the FIRST survey team present details of their radio-loud
BALs in Becker et al.\ 2000.  They include objects which don't satisfy
the accepted definition of BALs (Weymann et al.\ 1991) because their
absorption lines
are too narrow.  It's probably quite reasonable to discuss these
with the BALs, but one must not be misled regarding the incidence
of BALs among the radio loud objects.  Again many of the objects
in that paper are similar to OI287, and thus aren't new phenomena.

6.  BAL quasars show much higher optical polarization than average
for radio quiet objects (Schmidt and Hines 1999 and references therein;
Hutsemekers and Lamy 2000).
This is due at least in part to attenuation of
the direct ray, and so higher contrast for any scattered rays.
Similar behavior has been seen in far-IR-selected quasars.
Continuum polarizations usually
rise weakly to the blue, possibly because the direct, unpolarized
ray is reddened so the scattered ray has better contrast in the blue.
(The original IRAS quasar, 13349+2438, is qualitatively
similar: Wills et al.\ 1992.)

7.  The BAL clouds (or wind) generally lie outside the broad emission
line region since they partially absorb the emission lines.  
The emission lines don't share the continuum polarization.
Thus the high polarization must stem from scattering inside the emission
line region,
or cospatially or slightly outside it.  The lack of polarization of a broad
emission line of a BAL quasar 
was first shown for PHL5200, by Stockman et al.\ 1981.

8.  The troughs are almost always weaker in polarized flux than 
in total flux.  Probably some photons scatter around the absorbing
clouds.  It's not always true however:  take a look at the remarkable
behavior of FIRST 0840+3633 (Brotherton et al.\ 1997).  Both objects
in that paper show (rare) absorption from excited metastable states
in Fe II, and also show low-ionization lines in absorption
(``Iron lo-BALs").\footnote{Pat Ogle (pc) says the complex behavior of
the data on FIRST 0840+3633 can be explained by wavelength-dependent
dilution by unpolarized Fe II emission complexes.}
In this regard they are similar to only three
previously observed AGN (Halpern et al.\ 1996).
It would be fun to consider the radically varying polarization
behavior of the various transitions as functions of atomic parameters
such as critical density.  A similar object has been analyzed by
Lamy and Hutsemekers (2000).

9.  Taking a cue from the Unified Model, most workers suppose that
these are relatively high inclination quasars --- and that at
slightly higher inclinations they'd be classified as Ultraluminous
Infrared Galaxies with hidden quasars (see Sec 3).  

Nothing is known about the radio axes of most of these objects (if such exist).
The Classical Double BAL has a steep-spectrum core, and optical
polarization $\sim25$ degrees off from the overall radio axis (Gregg et al.\
2000).  It would be worthwhile to observe the core at high angular
resolution.

Finally, a clue regarding orientation can be gleaned from the radio
spectra.  Falcke and collaborators (e.g., Falcke et al.\ 1996)
have argued fairly pursuasively that flat spectra in radio
quiet quasars indicate a polar orientation, just as it does for
radio loud ones.  Perhaps the intermediate-inclination idea
just mentioned is favored by the mix of BAL radio spectral 
indices (Barvainis and Lonsdale 1997).

Summary:
  Many more thought-provoking details are available in the
reviews cited and referenced herein.  However the big-picture
consensus seems to be that the BALs may be at intermediate
angles, where the line of sight passes through a fast
wind atop a torus.

\section{Radio Galaxies}\label{sect:2}

\subsection{Unification with quasars}

\subsubsection{Optical properties}

This will be a quick history of the subject as I know it.
The usual apologies for omissions [actually, you can still tell me
before the printing!].  Relatively complete early references can be
found in my 1993 ARAA paper.  

\begin{figure} 
\centerline{\epsfig{file=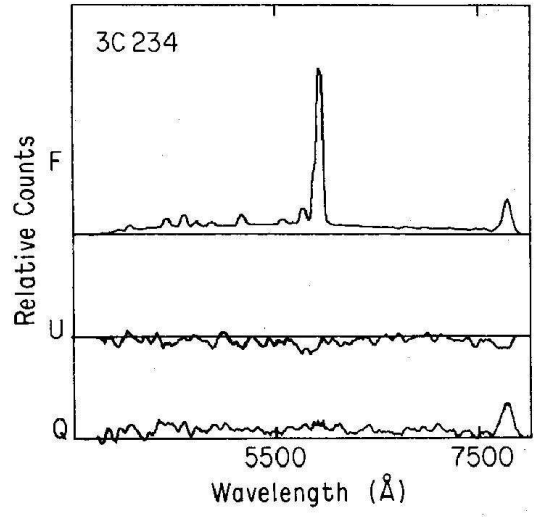,height=3in}}
\caption{The `polarized fluxes' for the Q and U Stokes parameters, as
originally     
discovered in 1982.  These represent the fractional Q and U spectra multiplied
by the total flux spectra.  For cases such as this, where the position
angle is pretty constant, these are like ordinary polarized flux plots,
but with a symmetric and unbiased error distribution.  Note that the strong
permitted line, H$\alpha$, shows in polarized flux and the forbidden [O III]
$\lambda$4959, 5007 do not.}
  \label{fig:3}
\end{figure}

\begin{figure}
\centerline{\epsfig{file=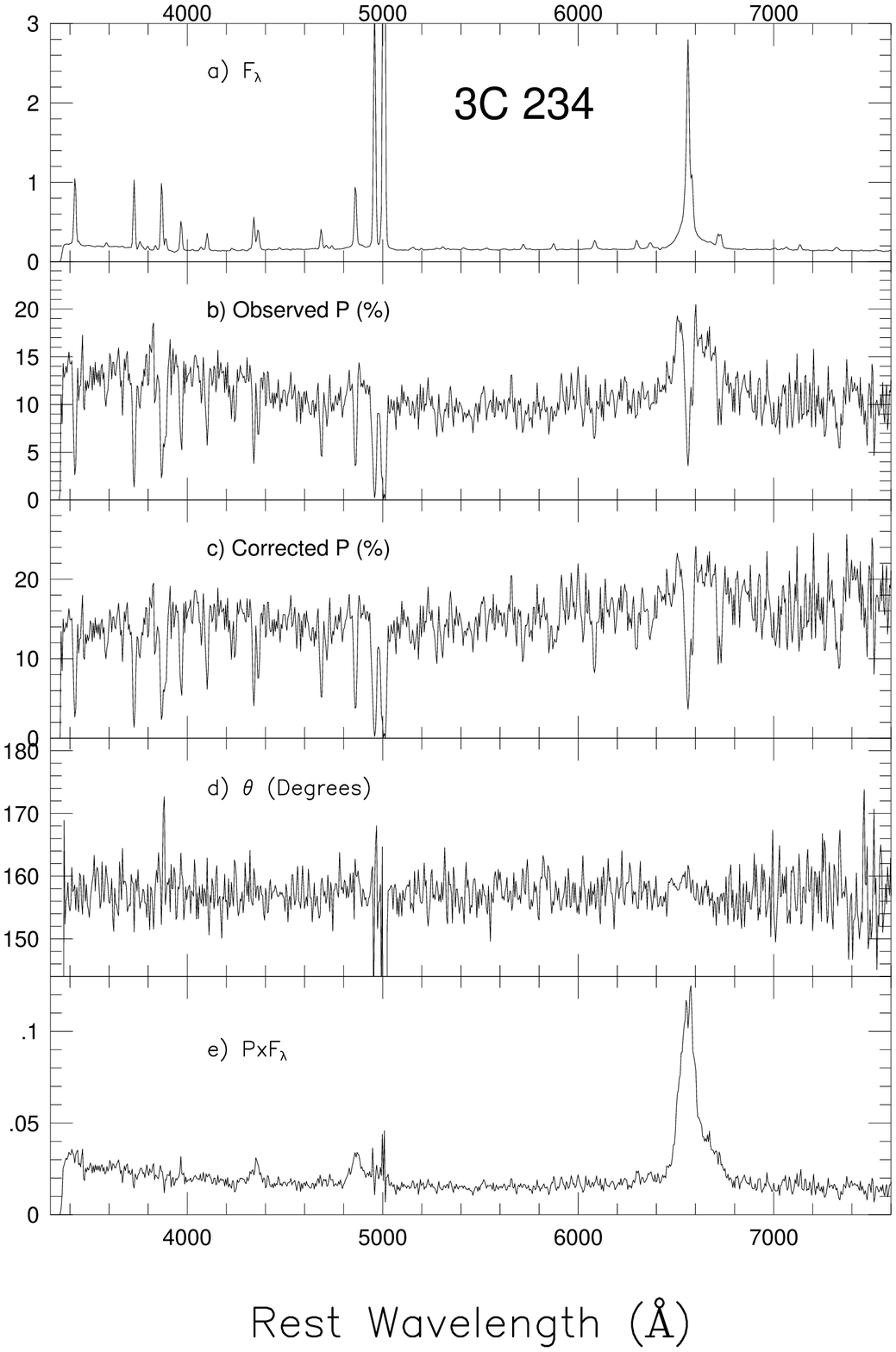,height=3.5in}}
  \caption{Modern quality spectropolarimetric data on 3C234 from Tran
{\it et al.}\/ 1995.}
  \label{fig:4}
\end{figure}

Most powerful radio sources have spectra in the same two 
spectral classes as for the radio quiet ones:  
Type 2 Radio Galaxies, usually called Narrow Line 
Radio Galaxies, and often just called radio galaxies if at high redshift;  
and Type 1, those with strong broad lines in
total flux which for historical reasons are called 
Broad Line Radio Galaxies or radio loud quasars, according
to luminosity, basically.  I show my original spectropolarimetric
data on the first recognized case of a hidden BLR (3C234)
in Fig.~\ref{fig:3}, along with modern Keck data (in a slightly
different form) by Tran {\it et al.\/} 1995 in Fig.~\ref{fig:4} (see
Kishimoto {\it et al.}\/ 2001 for the UV).
Excellent data and analysis of many more radio galaxies can be found in
Cohen {\it et al.}\/ 1999.

A great HST picture of a torus is that in 3C270 =
NGC 4261, found by \cite{Jaffe96}.
Note that spectropolarimetry provides
only the crudest information on the torus size.  I believe in 
most cases the inner radius is near the sublimation point ($\sim1$ pc),
because the spectral energy distributions show that all
Type 1 objects must have a substantial covering factor of
hot dust.  This HST image and some molecular mapping of various objects both
show a (quasi) outer edge at 100--300pc.

Radio loud AGN can also have a contribution (sometimes
dominant) from highly variable, highly polarized synchrotron 
radiation.  From the point of view of the radio-optical 
Spectral Energy Distribution, the optical synchrotron 
source is seen to be simply the high-frequency tail
of the radio core emission  (Landau {\it et al.}\/ 1986, Impey and
Neugebauer 1988).  These
objects are the blazars (Wolfe 1978).  Those with very low
equivalent width emission lines are historically called 
BL Lac Objects.  However, this subgroup has no physical
meaning, and since the objects vary, they alternate
their classification over time with this nomenclature!  
For example, BL Lac is 
often not a BL Lac object with this definition (\cite{Vermeulen95}).
Another problem with 
this nomenclature is that it mixes low-luminosity nearby
cases with FR1-level extended radio emission together with high
redshift very high luminosity  objects.  Currently some
people are saying that BL Lacs are specially oriented
FR1 radio galaxies.  This is sloppy in the extreme since 
a significant subset of them are known to be inconsistent with it
(e.g., \cite{Kollgaard92}).

I've often argued
that a better split for the blazars would be according to
whether their extended radio emission is consistent
with an FR1 or FR2 power level.  I think this has more hope 
of having physical meaning than setting an arbitrary
equivalent width limit on the emissions lines
on the discovery spectrum, and it would retain the
advantages of the latter for demographics.

Also, some authors have gone to the extreme of calling something
a BL Lac object largely because it doesn't have the strong
4000\AA\ break expected for late-type stars.  That seems crazy to me because it
includes zillions of faint blue starburst galaxies\dots
like the sources for the blue arc lenses in clusters.  
It took a long time even to get redshifts for the latter
because their spectra are so featureless.  High polarization
or at least high variability is required when defining
``BL Lac."

\begin{figure} 
\centerline{\epsfig{file=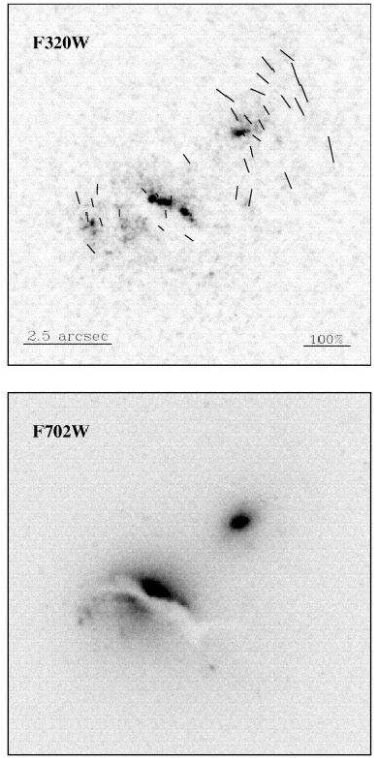,height=4in}}
  \caption{The spatially resolved ($\sim8$ kpc) mirror in 3C321, by HST
in the near UV (F320 FOC filter).
The lower plot shows the red (HST WFPC2 F702 filter) total flux image.}
  \label{fig:5}
\end{figure}

Next, let's consider some polarized images which resolve the
mirrors in Narrow Line Radio Galaxies.  Ideally, these are 
made in the rest-frame near-UV, where dust scattering 
cross-sections would be large, and where there is reduced
confusion with the host galaxy light.  Fig.~\ref{fig:5} shows our HST
image of 3C321 (\cite{Hurt99}).  
The pattern is centro-symmetric within
the noise, and P is locally rather large ($\ltwid50$\%).  
\cite{Young96} report scattered
broad lines from the hidden quasar.  Their data also show that
the polarized flux spectrum is indistinguishable from a total-flux
quasar spectrum, and this suggests unreddened electron
scattering.  It doesn't prove electron scattering though,
and we hope to look for spatially resolved scattered X-rays
to make a strong test.  It's very important to establish this with
certainty, because of the very remarkable consequences!

Many of the ($z\sim1$--2.5) ``aligned" radio galaxies (optical extension
parallel to radio) have behavior very similar to that at 3C321
(e.g., \cite{Best97}, and for the polarization, Cohen et al.\ 1999,
Dey et al.\ 1996, Cimatti et al.\ 1997, and Vernet et al.\ 2000).  In
order to have sufficient optical depth
in electron scattering, given the huge physical sizes of the 
polarized mirrors, the ionized gas masses are huge.  
Depending on assumptions, they may be $10^{12} M_\odot$ or considerably 
more.  Also, in cases where the broad lines are clearly seen in 
polarized flux, the putative scattering electrons cannot be
too hot --- at X-ray halo temperatures the scattering process
would broaden the lines beyond recognition.  
They must be $\ltwid10^6$K, which means they aren't 
typical cooling flows or hot haloes supported hydrostatically.
So all that mass would likely be falling in on a dynamical
timescale --- conceivably forming a central cluster galaxy
like M87.  A recent discussion can be found in Kishimoto {\it et al.}\/ 2001.

Incidentally, at even higher redshifts, the tiny sample
observed spectropolarimetrically so far does not show polarization, but instead
shows absorption lines probably (at least in part) from 
stellar photospheres.  These objects are necessarily
observed at shorter rest wavelengths ($\sim1500$\AA) so the comparison
with the $Z\sim1$ objects isn't completely clear.  It does show,
however, that almost everyone's theory for the optical radio alignment
effect is right somewhere (\cite{Dey97}).

\subsubsection{Radio properties}

There are two types of powerful extragalactic radio source,
the normal double sources and the core-dominant superluminal
sources.  (This oversimplifies, of course.)  It's rather well
established that this is mostly
another orientation effect:  the latter are seen in the 
jet direction, and are greatly boosted in flux because of
special-relativistic aberration.  Indeed, these core-dominant
sources generally show {\it radio}\/ ``haloes" consistent with lobes seen
in projection around the bright core, if mapped with good 
dynamic range (\cite{Browne82};  \cite{Antonucci85};
\cite{Wrobel90}, and many others).  

Now the double radio sources themselves can be divided into
morphological classes, by whether the lobes are edge-darkened
or edge brightened.  The former (``FR1") turn out to be 
basically the low luminosity objects.\footnote{For a very interesting
refinement of this statement, see \protect\cite{Owen94}.}  They have several
other correlated differences which gives this separation
some physical significance.  

I think that the {\it most luminous}\/ of the (luminous) FR2
are all just hidden quasars.  In many 
cases the polarized flux spectra already show it.  
In the 1980s, radio astronomers were making statistical
tests of the identification of double radio {\it quasars}\/ with
core-dominant superluminal sources and statistical
problems arose, such as finding ``too many" fast superluminals
relative to the expected beaming solid angle
(e.g., \cite{Laing88}, \cite{Barthel87}). These 
can all be explained by a dearth of objects with axes nearly
in the sky plane.  
The reason is now obvious:  quasars in the sky plane do exist,
but are called radio galaxies because the obscuring tori 
block our view of the nuclei.\footnote{Also note this interesting pair of papers
on the projected linear sizes: \protect\cite{Singal93};
\protect\cite{Gopal-Krishna96}.
They discuss whether the lower-luminosity FR2s
have unexpectedly small projected linear sizes,
relative to beam-model predictions.}

These very powerful radio galaxies have strong narrow emission
lines like Seyfert 2s, but many of the weaker FR2 types do not,
and {\it most}\/ of the (low-luminosity) FR1s do not.  Let's consider
first the FR2 radio galaxies:  are the relatively weak ones hidden quasars?
The optical-UV continuum is thought by many people to be
thermal radiation from optically thick matter falling into
a black hole.  If there is instrinsically no hidden quasar (BBB),
then according to current theory, there can't be much of
an accretion flow.  In that case only the black hole spin energy 
would be available to produce and sustain the powerful radio jets.
These objects can be called nonthermal AGN (if they exist).  
%It does seem though that at least a few radio galaxies with low-ionization
%relatively weak narrow lines {\it do}\/ have pretty good evidence
%for a hidden quasar (see \cite{Sambruna00} for Hydra A). 

``Optically dull" FR2 radio galaxies show very little optical
polarized light in general.  The reason could be:  there's
no hidden quasar;  there's no mirror; the hidden quasar
is completely surrounded by dust; relatively low-column kpc-scale
foreground dust lanes block our view of the scattered nuclear light
as in 3C223.1 (Antonucci \& Barvainis 1990).  A more robust test for 
a hidden quasar would be looking for the inevitable ``waste heat"
in the mid infrared.  D.~Whysong are I are trying to make this
test at Keck, but it's pretty hard.  (A few relevant objects
were observed by ISO.)

Now let's consider the FR1 sources:  {\it most}\/ of these are optically
dull:  the narrow line emission is very weak, and of low
excitation.  This isn't suggestive of a hidden quasar, but again
the inner narrow line region could be occulted, or
the nucleus could be completely surrounded by absorbing dust.  In a
very significant series of papers, Chiaberge {\it et al.}\/ (1999, 2000),
show that archival HST images have nuclear point sources in {\it most}\/ of the
optically dull 3C galaxies of both FR types.  They reason that an AGN cannot
be hidden in most optically dull radio galaxies because we can seemingly see
into the center in order to detect the point sources (thought by the authors
to be synchrotron emission from the tail of the radio core spectra).  This
behavior is unlike that of Seyfert 2s and the very powerful narrow line radio
galaxies, and ultraluminous infrared galaxies, all of which show
{\it no}\/ point source in the optical band.\footnote{At least a couple of
optically dull radio galaxies with pointlike nuclei
do have strong high IRAS fluxes, arguing for hidden nuclei in those cases.}

A good example of a somewhat optically dull FR1 (or FR
borderline) radio galaxy is M87 (\cite{Reynolds96}).  It shows
a nuclear point source in the optical, and has no powerful thermal IR source.
We find that at 11.7 microns there is only a weak $\sim15$mJy 
point source, and this could easily be
explained as nonthermal emission from the base of the jet.
{\it M87 really can't have a hidden AGN with luminosity remotely comparable
to the jet power}\/ (Owen {\it et al.\/} 2000).

On the other hand, the nearest FR1 is Centaurus A\dots which 
has {\it no}\/ optical point source, lots of thermal dust emission, and 
considerable polarimetric evidence for a hidden ``thermal" optical/UV
nucleus (\cite{Marconi00}, \cite{Capetti00}).
Thus the FR1 family is a 
heterogeneous one.  Also regarding Cen A, it's well worth 
taking a look at the
beautiful CO 2-1 torus image, Fig.~2 of \cite{Rydbeck93}!

%Is it possible for optically dull radio galaxies to have a 
%totally absorbed optical nucleus (unit covering
%factor as seen from the nucleus)?
%Remember also that 

The FR1 radio galaxy
3C218 (Hydra A) has (relatively) weak emission lines
of low excitation (\cite{Ekers83}), yet strong 
evidence for a hidden AGN (\cite{Sambruna00});
it shows that nuclei can have hidden AGN even if they are
rather ``optically dull''.

A related question is whether a quasar
or Broad Line Radio Galaxy can have an FR1
radio source.  {\it A few are in fact known}\/ (e.g., see \cite{Lara99}).
As those authors put it,
this makes their source ``a nontrivial object from the point
of view of current unification schemes."  Another recent case
is from \cite{Sarazin99}:  1028+313.

\subsubsection{The range of optical slopes for lobe dominant radio quasars}

There is one more issue I want to discuss, and it's actually
terribly important.  A major study of the Unified Model,
based on a well-selected sample, was presented by Baker (1997).
That paper finds a weak correlation between radio core dominance
(thought to be a crude statistical inclination measure)
and optical slope.  The sense is that those with weak cores
are redder.  Furthermore for a subsample, the broad line
Ba decrement is steeper at low core dominance.  And the loop is
closed by the finding that the slope is correlated with the
decrement in the consistent sense.  
Baker concludes from all this that the slopes are mainly 
influenced by reddening, and that reddening decreases with the
``latitude" of the line of sight.  The only obvious weakness here
is the poor quality of the core-dominance/slope correlation.

This conclusion that the optical slopes are set by various amounts
of reddening is a profound one.  First it shows that, even among
lobe-dominant quasars (thought to be seen from the general polar direction,
like Type 1 Seyferts), the great majority of the luminosity
is absorbed by dust (Baker's Fig.~15).  Second, it encourages
theorists to dismiss the steeper slopes, and just make models that
can fit the bluest ones.  The latter is crucial for accretion disk
models!

\begin{figure} 
\centerline{\epsfig{file=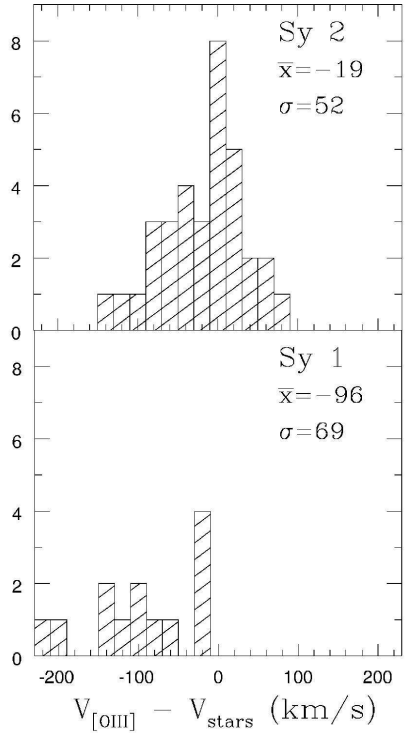,height=4in}}
  \caption{Central velocities of the narrow line region in the two types
of Seyfert.  Plot from Todd Hurt, using data from Mark Whittle.}
  \label{fig:baker}
\end{figure}

Core dominance is only a crude indication of orientation because
1)  the core relates to the luminosity averaged over a timescale
of years, whereas the lobe emission averages over many millions
of years;  and 2) the lobe luminosity is quite dependent on the
environment on $\gtwid10$kpc scales.  Wills and Brotherton tried
normalizing the core fluxes to the optical fluxes instead of those
of the lobes.  This applies to radio quasars and broad lines
radio galaxies only, and is considered undefined for blazars
and narrow line radio galaxies.  While there's no a priori
reason to expect this to work well, it certainly seems to!  That
is, several previous correlations of ostensibly orientation-dependent
parameters using the old core dominance, look much better with this
new version!  

D.~Whysong (pc 2000) tried Baker's plot of core dominance vs optical slope
with the new parameter --- and the plot looks even worse than the original.
This made us reconsider the reddening interpretation. We don't
think foreground reddening can make the steep slopes, because there
is no exponenetial cutoff in the UV.  Quantitatively this is compelling,
and just requires that the extinction rise with frequency, as needed
also for the reddening explanation.  The only obvious way around this
is to have the continuum emitter and the dust cospatial and intermixed.
That would be quite amazing for the Big Blue Bump.

I suggest that the slopes are actually intrinsic, and that the
correlation with Balmer decrement is an ionization-parameter effect.
Certainly the steeper ones have a small fraction of the ionizing
photons of the flatter ones, for the same strength of the optical
emission lines.  It doesn't correlate with orientation
at all in this picture.  More cogitation is required.

\subsection{Observing the ionized intergalactic medium}

Radio polarization maps of high-redshift radio galaxies and quasars
could supply a key cosmological parameter.  We've been trying to detect
Thompson-scattered radio halos to detect the ionized intergalactic medium.
This was originally suggested by Sholomitskii.
The expected baryon density based on nucleosynthesis is $\sim5$\% of closure,
corresponding to IGM optical depths of $\sim10^{-3}$ over the expected halo
size ($\sim10^{7-8}$ light years, based on guessing the AGN lifetime).
Sensitivity is needed on large scales 
and it turns out the Australia Telescope Compact Array
is the best choice.

Our first attempt at this was published in Geller {\it et al.}\/ 2000,
providing an upper limit to the IGM density
below 100\% of closure (you have to start somewhere).
We hope to do much better.  If we can eventually detect the halos
with confidence, we'll learn the quasar lifetime and beaming pattern, too,
in principle.

\section{Ultraluminous Infrared Galaxies}\label{sect:3}

\subsection{Spectropolarimetry and Type 2 quasars}

\subsubsection{Quasars of Type 2}

\begin{figure} 
  \centerline{\epsfig{file=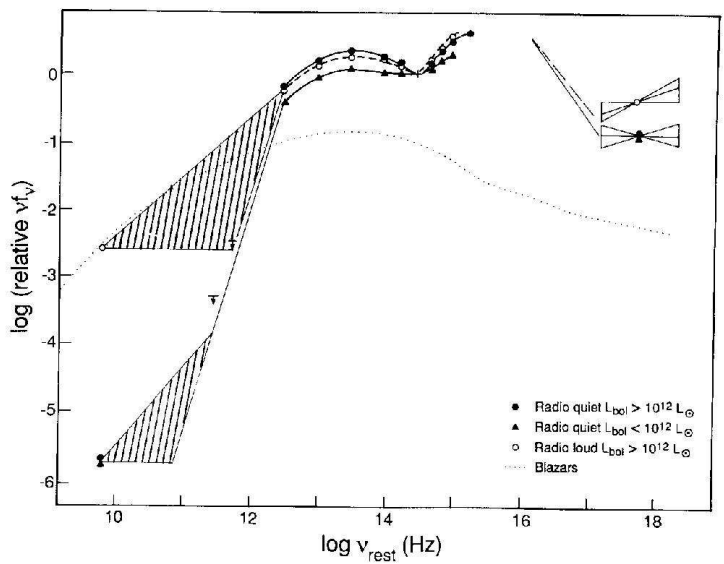,height=3in}}
  \caption{Composite spectral energy distribution for AGN (Sanders
{\it et al.}\/ 1989).  Note: the continuity between far IR and radio
is not correct for most objects, but this plot is valid in the energetically
important parts.} 
  \label{fig:sed}
\end{figure}

Let's look again at the spectral energy distribution of unobscured AGN
(Fig.~\ref{fig:sed}).  Obscured AGN are similar except the optical/UV
is much lower,
and sometimes the X-rays are as well.  What is your definition of a 
``Quasar 2"?  I'd expect that a Seyfert 2 with an extremely powerful
nucleus would have an extremely powerful infrared bump, absorbing
the AGN light, and reradiating it as in relatively isotropic infrared dust
emission.

Next I'd expect a Quasar 2 to have powerful high excitation
narrow emission lines, since the Narrow Line Region is generally outside
the torus, and hence it emits rather isotropically.  It would not scale
with the infrared, simply because empirically the narrow line equivalent
widths decrease with luminosity in the objects seen directly (Seyfert 1s
and radio quiet quasars).  But it should be much more powerful than
those in the Seyfert 2s.

I'd also expect, based on the Seyfert 2s and the unified model, that
there would be {\it no}\/ optical/UV point source, and that the hidden Type 1
nucleus would appear at good contrast in the polarized flux spectra.

Note that the optical/UV continuum flux would {\it not}\/ scale with the AGN
luminosity, because that continuum in Seyfert 2s is almost always
strongly dominated by the light from the host galaxy.  Suppose 90\% of the 
Seyfert 2 continuum derives from light from the underlying old
steller population, a fraction not in great dispute.  Then a simple scaling
of the AGN power by a factor of 10 would lead to an increase in optical/UV
flux of only a factor of two.  This is oversimplified given differences among
the AGN nuclear regions (reddening, young stars), but I think it's
qualitatively correct.

I've just described {\it exactly}\/ the high-ionization ultraluminous infrared
galaxies (e.g., Hines et al.\ 1999).  I still read that there are no
Quasar 2s, or that their
existence remains to be demonstrated.  If you think that this class remains
undetected, please tell me your definition of Quasar 2.

On the radio side, I think there would be little argument that the Type 2s
are the powerful narrow line radio galaxies --- at least for those with
the highest radio luminosity.

\subsubsection{Properties of Quasars of Type 2}

There are a {\it lot}\/ of high-ionization ultraluminous infrared galaxies.
The estimates I've seen indicate that the fraction of all infrared
galaxies comprised by those of high excitation rises with infrared luminosity,
and reaches half of those more luminous than 1--$3\times10^{12}$
Lo.

Fig.~\ref{fig:sed} shows the generic SED shape
for unreddened quasars.  It is important that the IR bump has about 30\%
of the integrated flux.  The simple and plausible interpretation is that
the ``tori" cover $\sim30$\% of the sky, as seen from the nuclei.  This
seems at first glance to be consistent
with the space density of high ionization ultraluminous galaxies relative to
quasars matched in apparent bolometric flux (\cite{Gopal-Krishna98}).
However, the correct thing to do here is to compare ULIRG and quasar
space densities {\it as matched by far-IR, since that, not L$_{BOL}$, is
isotropic}.  We can conclude that the (UV-selected) far PG quasars are
well below average in dust coverage, as deduced by many others.

\begin{figure} 
  \centerline{\epsfig{file=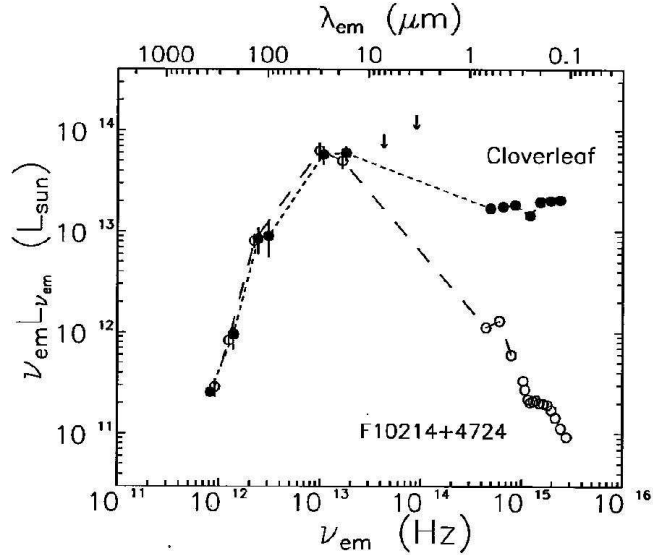,height=3in}}
  \caption{Compared with an unobscured quasar,
the Cloverleaf is depressed and reddened in the Big Blue Bump region.
For the infrared galaxy (ULIRG) {\it all}\/ of the direct nuclear photons are
absorbed.} 
  \label{fig:hyperclover}
\end{figure}

Fig.~\ref{fig:hyperclover} shows a comparison of the SED between the first
``hyperluminous"
infrared galaxy (F10214+4724), and the ``Cloverleaf" reddened quasar.  Both
are lensed and the quantitative close agreement in the far-IR is fortuitous.
But it makes the point that even a somewhat-reddened quasar has a much
higher fraction of the light in optical/UV.  The infrared galaxy has no 
point source visible in those regions, but there is a little scattered
light from the hidden quasar which makes it detectable in the optical/UV.
I say this with confidence because the hidden quasar appears in the polarized
flux plot, just as in nearby Seyfert 2s  (Goodrich {\it et al.}\/ 1996).

It's also known that the relatively isotropic CO/far-IR ratios are generally
the same in quasars as in ultraluminous infrared galaxies, even for those
classified as starbursts (Alloin {\it et al.}\/ 1992;
Evans {\it et al.}\/ 2001).

The IR-warm ultraluminous infrared galaxies start to reveal a pointlike
nuclear source in the near-IR (e.g., Surace, Sanders \& Evans 2000).
These are attributed to penetration of the outer parts of the absorber
well enough to see the very warm unresolved nuclear dust.  This hot dust
can be reached much more easily, apparently, than the nuclei themselves.

Now an extremely important question arises, in terms of understanding
the global SEDs.  What is the torus column density and the opacity at various
wavelengths?  Starting with NGC 1068, the lack of a strong variable X-ray
source immediately shows that many are ``Compton thick," blocking even the
$\gtwid10$keV photons.  At solar abundances this means the column density is
$\gtwid10^{25}$ cm$^{-2}$.  Recently it's been possible to measure the
column density (or limit) distribution for fairly complete Seyfert 2
samples and the median is almost that high (Maiolino {\it et al.}\/ 1998;
see also Salvati and Maiolino
1988).  It immediately follows that in most cases 
%we {\it cannot}\/ probe the nuclear regions in the mid-IR.  S
{\it mid-IR observations reveal the conditions some fraction
of the way into the tori unless the dust/gas ratio is extremely low}.
Apparently, there is star formation in the tori,
since the mid-IR observers often see ``starburst" spectra.  {\it Whatever
the dust-gas ratio, the point is that the Type I nucleus is seen in the
x-ray but not in the mid-IR.}

\subsection{Distinguishing hidden AGN from hidden starbursts in ULIRGs}

\subsubsection{Opacities and Luminosities}

There has been a great deal written on this subject which implicitly or
explicitly presumes the tori are optically thin in the mid-IR.  I can't
put much faith in these papers because we have very strong evidence that
this would not be the case in typical Seyferts.  Thus the spectra
that simply show starburst mid-IR lines are probably just studying 
conditions inside the tori.
%A convincing argument for a dominant starburst would be certain emission
%lines indicate
I'd be convinced of an important starburst, if some observed spectral
features indicated {\it starburst luminosities}\/ consistent with the
far-IR power.
However, even in such cases I reject the claim that an object or certain
objects get ``most" of their energy from starbursts.  That language
requires the starburst energy contribution to be $75\pm25$\% of the total
luminosity.  But there is no such precise bolometric luminosity
that can be inferred from any emission line feature.  Similarly, the best
predictor of hidden AGN luminosity is the hard X-rays --- for objects
in which they get through the torus.  It's pretty robust but only works
at the factor-of-three level, certainly not sufficient for anyone to
say the hidden-AGN luminosity is $75\pm25$\% of the total.  As soon as 
somone makes a statement about energy sources with this kind of precision
I tend to stop reading because I've lost faith in my author.

Join my new group:  the Militant Agnostics.  Our motto is, ``I don't
know, and you don't either!"

\subsubsection{What are the LINER ULIRGs?}

My {\it guess}\/ would be that the ULIRGs with high ionization narrow lines
are mostly powered by AGN, and likewise for the starbursts.
  It's just a guess.  But what about the large minority of ULIRGS which 
have LINER (Low Ionization Nuclear Emission Region) spectra?  
Here we can't even tell if the region producing
the lines {\it that we see}\/ derives from a hidden AGN or a starburst.

A remarkable paper and those surrounding it illustrate the point about
LINER ULIRGs.  After extensive studies of the mid-IR spectra of ULIRGs,
it was claimed by many that their energy source is usually a hidden starburst.
Then some mid-IR experts got together with an optical AGN spectroscopist,
in part to see what the mid-IR spectra have to say about the optical LINERs
(Lutz, Veilleux and Genzel 1999).
They showed in a remarkably clean manner that those with optical starburst
spectra also have infrared
starburst spectra;  same for the AGN;  but the interesting part is that
the many optical LINERs show starburst infrared spectra!
This was interpreted
as indicating that the LINERs are simply a slightly different manifestation
of a hidden energetically dominant starburst.  Only trouble is, I think
few if any of these mid-IR spectra came from the actual nuclei, as discussed
above.
Maybe the observations prove that there is a lot of star formation
inside the tori, but not that it contributes $75\pm25$\% of the luminosity.

A good example is NGC6240, a famous ``prototype" (see Table 1 of Lutz
{\it et al.}\/ 1999).
The optical spectral type is listed as LINER, and the mid-IR is listed as
starburst.  (It is even a ``prototype" starburst in Genzel {\it et al.}\/ 1998.)
The simplest conclusion is that it's simply a starburst galaxy, with little
AGN contribution.  Well, the X-ray opacity at $\gtwid10$ keV is a few times
smaller than that in the mid-IR, so for a small (but non-negligible)
number of cases the X-ray penetrates, though the mid-IR doesn't.  Shortly
after (or perhaps contemporaneously with) the Lutz {\it et al.}\/ paper,
Vignati {\it et al.}\/ (1999)
published the BeppoSAX spectrum, covering the high energy as well as the low
energy X-rays.  It clearly shows a column of $\sim2\times 10^{24}$ cm$^{-2}$,
so that there would indeed very probably be high opacity in the mid-IR.

The hard X-rays are strong (direct), penetrating and rapidly variable,
and really must come from a hidden AGN.  The corresponding bolometric AGN
luminosity is to within uncertanties just that observed.  
Even in that case, given the factor of $\sim3$ dispersion in the
X-rays/bolometric ratios seen in other AGN, it would be too much to claim the
AGN luminosity is $75\pm25$\% of bolometric.  But it sure isn't negligible
either, so the mid-IR spectrum is {\it not convincing evidence
for a dominant starburst for this object and thus potentially for any of them.}

\begin{figure} 
  \centerline{\epsfig{file=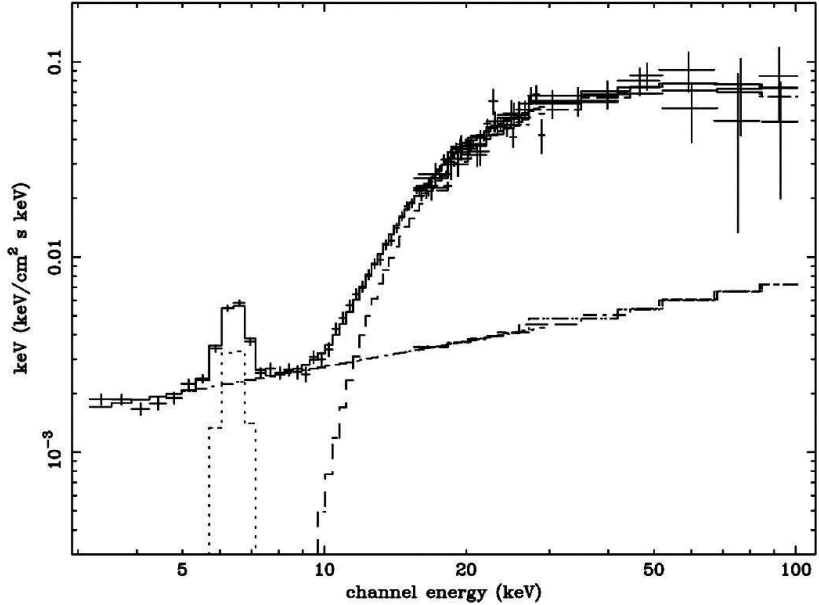,height=3.25in}}
  \caption{Wideband view of a Seyfert 2 X-ray spectrum with a column density of
$\sim5 \times 10^{24}$ cm$^{-2}$.  The object is NGC4945.} 
  \label{fig:9}
\end{figure}

There are many similar cases, but this is getting far
afield from polarimetry.  I'll just cite and show the case of NGC 4945.
It also is a ``template" starburst in Genzel {\it et al.}\/ 1998.  Yet it has
powerful rapidly-variable $\sim10$keV flux, and certainly contains a powerful
hidden AGN.  The reason I bring this one up is that it has a wonderful
wide-band X-ray spectrum, and the soft and midrange absorption indicating
the very high column couldn't be
clearer (Fig.~\ref{fig:9}, from Madejski {\it et al.}\/ 2000; see also
Eracleus {\it et al.}\/ 2001 for another example.).

The mid-IR observers aren't convinced though.  In a ``reply" paper they say
that ``the starburst may well power the entire bolometric luminosity" of
NGC4945
with the caveat that they can't really prove that an AGN doesn't provide
``up to 50\% of the power''.  Don't you love science?

The recently-discovered ``Scuba", millimeter-selected
galaxies are quite similar.  They can be detected easily because of a
very favorable
``K-correction" for objects at significant redshift.  We probably know a tenth
as much about these compared with the old ULIRGs.  Yet at least one of the
discovery papers simply {\it assumes}\/ they're all powered by star formation,
and doesn't even mention the AGN possibility.  This unjustified assumption
has major consequences for the luminosity density history of the universe
(Madau Diagram),
for the X-ray background (Almaini {\it et al.}\/ 1999), and also for black hole
demographics.

\section{Emission Mechanism for the Big Blue Bump Spectral Component}\label{sect:4}

\subsection{Introduction and Polarization Diagnostics}

We have some hope of understanding quasars because the spectral
energy distributions are quite generic.  Aside from an occasional blazar
(radio core, synchrotron) component, they seem to comprise an 
IR bump (thought by almost everyone to be thermal dust emission), a usually 
energetically
dominant ``Big Blue Bump" optical/UV continuum component, and an
extremely interesting but less powerful X-ray component.  Refer again to
Fig.~\ref{fig:sed}.  Note that the X-ray component does
become
competitive with the Big Blue Bump (BBB) in some of the lowest-luminosity
nuclei.

Since the BBB is generally energetically dominant, and peaks to order
of magnitude in the right spectral region, it is often assumed
to be thermal emission from optically thick accreting matter.  Published models
almost all assume the emitter is a standard thin opaque, quasistatic
disk converting gravitational potential energy through viscosity
into radiation.  These are sometimes called ``Shakura-Sunyaev" disks.
Almost no one would argue that this is qualitatively correct any more.
Nevertheless most theorists assume {\it de facto}\/ that whatever is
going on, you get the same spectrum!

Polarization has played a role in testing the accretion disk model as
follows.  The tale is a bit convoluted.  In an optically-thick disk
with a scattering atmosphere and
with the heat deposited at large optical depths, the polarization
should range from 0 to 11.7\%, and always lie along 
the (projected) disk plane.  However, it's been known since
the world began that this doesn't describe the observations. 

The polarization degree 
might be okay if the edge-on quasars were missing (manifesting as
NLRGs or Infrared Galaxies).
There are claims that the polarization magnitude distribution
actually is consistent with the predicted one, without the
high-inclination objects.  However, in the case of lobe dominant,
steep spectrum radio quasars, the direction is observed to be
{\it parallel}\/ to the radio jets (\cite{Stockman79}).
%If the sign is wrong, getting the
%magnitude right doesn't help when balancing your check book
%(J Miller, pc), and it doesn't help here either.
%
Thus these early papers implicitly assumed that the jets emerged from
the sides of the disks.  A possible way out is to claim that
for these radio-loud objects, the polarization derives from
an optical blazar component, but this is observationally
untenable because the radio core spectra decline sharply by
the millimeter region (e.g., Antonucci et al 1990, Knapp and Patten 1991
and van Bemmel \& Bartoldi 2001).  Also as noted above, Seyfert 1 galaxies
probably show the same effect, and they have very weak radio cores.

In the models of Laor, Netzer and Piran (ca.~1990 vintage), absorption opacity
simply dilutes the effects of scattering in terms of magnitudes	of
polarization, resulting in a strong rise in P with frequency
followed by a sharp decline past the Lyman edge.   
\cite{Antonucci96a} and \cite{Koratkar98}
tried to test the polarization predictions (low in the
optical, higher in the UV longward of the Ly edge).
No object seems to have the
expected behavior.  In fact several showed very high polarization,
only below $\sim750$\AA\ in the rest frames (\cite{Koratkar95}).
Regarding the latter, I'm reminded of an apochryphal Eddington
quote:  never trust an observation until it's
confirmed by theory.

It was discovered later (actually rediscovered, e.g., 
Gnedin and Silant'ev 1978 and probably earlier ones;
some modern papers are \cite{Matt93}, Blaes and Agol 1996) that Laor's
seemingly reasonable way of accounting
for the effects of true absorption is really not correct in
most of parameter space.  The current state-of-the-art
as far as I know is Blaes and Agol 1996.  They can
get a slight parallel polarization, at least in models that
produce too-few ionizing photons, {\it modulo}\/ Comptonization.  
%But we still see no great resemblence to the observations.
But the latest wisdom says the disks will be completely Faraday-depolarized
(Agol and Blaes 1996; Blandford, this volume).  Perhaps the observed
``parallel'' polarization is impressed downstream (though inside the BLR).

Getting the AGN geometry right at {\it any}\/ distance from ground-zero
would be valuable, and \cite{Chen90} made some very interesting,
very specific predictions for the polarization behavior in
double-humped broad H-$\alpha$ profiles --- the latter were argued
to arise in a thin disk at a larger radius than the BBB.  While
these ``smoking gun" predictions weren't confirmed (\cite{Antonucci96b},
\cite{Corbett98}) it was later shown that a range of polarization
behavior could occur in the disk model, if certain assumptions
were relaxed (\cite{Chen97}).

\subsection{Other Diagnostics}

I'll make brief mention of some other tests of the accretion disk
paradigm for completeness.  My point of view is spelled out in
more detail in \cite{Antonucci99}.

\subsubsection{Lyman edge}

Most thermal models predict observable changes in the spectra at the
position of the Ly edge.  There are certain rather narrow regions
of parameter space in which they are not present at a detectable level.

The Shakura-Sunyaev disk generally leads to an edge in absorption,
as in stellar atmospheres in which this spectral region is energetically
important.  Kolykhalov \& Sunyaev (1984) considered this in some
detail.  No such edges are seen, however (Antonucci et al 1989)\footnote{Small
features have been claimed in composites (e.g., Zheng {\it et al.}\/ 1998,
but see \protect\cite{Tytler93}), but I don't think credibly in individual
objects (see \protect\cite{Appenzeller98} for a great observation of 3C273).
Since the absorption-line test described above probably wouldn't work well
in composites, it's not clear that such a feature is
due to a disk edge.  At least that's the lesson Kinney and I learned from
our follow-up observations of disk edge candidates.}.  The key test
to determine whether or not a candidate edge really comes from 
a relativistic disk rather than foreground material
is a {\it lack}\/ of an accompanying set of sharp absorption
lines:  A.~Kinney and I once thought we had some candidate disk edges but they 
all failed this test.  {\it Often people attribute an observed edge
to an accretion disk without bothering with or even mentioning
this basic test.}

Later authors pointed out that Kolykhalov and Sunyaev were unable to
consider surface gravities below that expected for supergiant
stars, and that lower gravities and so densities 
were quite reasonable and result in smaller
edges.  It's amusing that a surface gravity and so density
much {\it higher}\/ than previously contemplated might also do the trick:
(Rozanska {\it et al.}\/ 1989, but see also Czerny  \& Pojmanski 1990
and Czerny \& Zbyszewska 1991).
Also the range of gravitational redshift of the emitting
elements and (for high-inclination disks) the range of Doppler
shifts would tend to smear the edges out.  For certain parameters they would be
very hard to detect.  Invoking too much inclination would
require reconsideration of the polarization distribution, and of consistency
with the Unified Models.

It is very interesting to consider now the flux and polarization behavior
at the Ba edge.  In some quasars the BBB polarization is $\sim1$\%, and
wavelength-independent.  More importantly, it arises inside the BLR
because neither the broad emission lines nor the ``Small Blue Bump"
atomic emission in the 2500--3500\AA\ region are polarized.  Thus the polarized
flux plot is a wonderful way to scrape off the atomic emission and see what
the BBB is doing in this region, as mentioned in Sec.~\ref{sect:1}.  So
far it looks like there is
{\it no}\/ Ba feature in the BBB (\cite{Antonucci88};  \cite{Schmidt00}).
Explanations of the lack of any Ly edge feature which depend upon
relativistic effects, or other effects occuring in the hot innermost
annuli, would not work here.  In some cases as noted below, the disk
continuum in the optical region must match rather red observed spectra
(spectral index $\ltwid -0.5$).  (The negative slope isn't a results of
foreground reddening since we see no downturn in the UV.)  {\it Models
must achieve this without producing a Ba edge feature.}

Much stricter constraints on, or detections of, a Ba edge feature would
be a very worthwhile spectropolarimetric project.  (I haven't been
able to get telescope time to do it!)

%I often complain that theorists will make models which address
%an individual problem with a more careful calculation or an ``epicycle,"
%in a way which is inconsistent on other grounds.  However, a good
%reply is that we critics can't combine constraints from different
%objects.

\subsubsection{Spectral energy distribution}

Early accretion disk models predicted positive spectral indices, well
longward of an exponential cutoff, whereas almost all quasars and AGN
have negative spectral indices (two good studies of the latter are
\cite{Neugebauer87} and \cite{Francis96}).  The observations weren't fitted
to the model optical/UV directly, but only after subtraction
of an ``infrared power law," extrapolated under the optical (e.g.,
Malkan 1983).  Indeed, \cite{Laor90} states that such a power law
is required.  Only trouble is, everyone is now convinced that the
IR is dust emission, which must drop like a stone at $1\mu$ and cannot
legitimately be extrapolated under the optical (e.g., \cite{Barvainis92}
and references therein).  It isn't yet clear whether pure disk models can
fit the optical observations (including the lack of Ba edge).
%Some models were later published
%that did manage to make such relatively red optical continua.

Although the dependence of disk maximum temperature on luminosity
(for a given $L/L_{\rm Edd}$) is only to the 1/4 power in the standard
disk, that's enough to predict dramatic differences between the
turnover frequencies of high luminosity quasars compared with low
luminosity Seyferts.  No such difference is found (e.g., \cite{Walter93};
also Mineo {\it et al.}\/ 2000 show that quasars have a steep rise
below 500eV just like Seyferts, na\"\i vely at least suggesting a similar
temperature).  In fact in the optical range people have typically
found that the more luminous AGN are {\it flatter}\/ (hotter):  see e.g.,
Fig.~5 from \cite{Mushotzky89}.

A more robust prediction is that for a given object, the fitted
temperature should vary as the brightness changes.  This is seen
{\it qualitatively}\/ in the UV for most objects.  In fact they all
behave this way, except the ones that don't (e.g., Antonucci 1999).

In a particular case, the extremely
luminous quasar HS 1700+6416, the spectrum extends to far too high
a frequency for the ``standard" model (\cite{Reimers89}), though
hard photons could always come from Comptonization (Siemiginowska
\& Dobrzycki 1990).
I've noticed also lately that it's become
socially acceptable to assume that the inner edge of the optically
thick disk can be anywhere needed to help fit a model.  Previously,
attempts were made to fit disks which extend into the last
stable circular orbit.  This may be reasonable physically, but
it conforms to the pattern that a new parameter is adopted for
each new observational fact.  Certainly the disk model has shown
no predictive power.

Finally there is the interesting issue of microlensing variations.
Rauch \& Blandford (1991) showed that for the Einstein cross
{\it any}\/ optically thick thermal
model which can produce the optical-region SED must have a thermodynamic
emissivity much greater than 1!  Others have disagreed e.g., \cite{Czerny94},
%and several others
on the grounds that 1) everyone knows a plain disk doesn't fit the
optical slope anyway, so we can invoke another component there from outside
the microlensed region, and 2) the observed ``caustic crossing" may
have been a rare event.  I like the Rauch and Blandford
color-temperature constraint because a disk could, in fact, produce the
observed SEDs given a certain heating of the outer annuli by the
inner ones, but on the other hand, recent data suggest that the
variation analyzed was, in fact, unusual and is legitimately modeled as a
2--$3\sigma$ event.

\subsubsection{Variability}

As far as I know, \cite{Alloin85} should get the most credit for
pointing out explicitly that AGN variations are much, much too fast
and too phase-coherent with respect to wavelength for any quasistatic
model.  Almost equally important in this context is the fine
quasar spectral variability study by \cite{Cutri85}.  Somehow these
papers didn't sink in for a decade or so, with claims being made that
the problem was first discovered by a much later 
monitoring campaign on NGC 5548!
The latter did provide the best limits on any lag between the long-
and short-wavelength variations which was so tight as to require
communication between the two relevant annuli at of order light speed
(\cite{Krolik91})!  This is a profound fact that shouldn't be ignored.
In the disk models, this is much shorter than the sound crossing time
as well as the viscous time --- for disk models it really requires
something tapping into the basic energy source with speeds of order
the speed of light!
This broadband variability, and the zero spectral index of the variable
part of the spectra, are more consistent in principle with hot but
optically thin emission (\cite{Barvainis93}).

Excellent recent data on variability can be found in \cite{Giveon99}.
The rates of flux variations may surprise you.

The rapid in-phase variability has led to the speculation that the
energy is actually dissipated in a ``corona," thought also to produce
the X-rays in the $\sim5$keV region.  Then photons from the corona could heat
the disk, rather than internal dissipation.  This is a real
``non-starter" since it's energetically untenable for all objects except
a few at the lowest luminosities.  
Remember that the rapid in-phase variability is a generic problem,
known at least since 1985 to be applicable to luminous quasars as well
as Seyferts.  See 
%the sketches in Koratkar and Blaes 1998, reproduced here as
Fig.~\ref{fig:sed} for a clear picture of these SEDs.

Although the energetics is totally damning for heating by the observed
X-ray source, I'll also mention that the
5-keV X-ray and optical continua do not vary together as expected in any
object, though ``second-order" predictions (or postdictions) are somewhat
as expected in at least one object, according to
Nandra {\it et al.}\/ (1998, 2000).  There sure doesn't
seem to be any near-IR vs.\ X-ray relationship (\cite{Done90}).  
Also a prediction of an unseen Ly edge in emission might
be a problem with external illumination of a disk (\cite{Sincell97}).

The X-ray Fe K-$\alpha$ profiles look like they could come from Kerr disks
and I thought showed at least that there are passive disks present in AGN.
However,
the line (and ``Compton hump") variability is virtually inexplicable in
that or any other picture; see \cite{Weaver00} for a brief review.

I can't think of any accretion-disk predictions that have come true, as far
as producing the BBB is concerned.  I don't even think the passive-disk
predictions for the K-$\alpha$ line count for much until its variability
properties can be reconciled with the disk picture.  As Vince Lombardi
might have said, in science prediction isn't everything.  It's the only thing.

\section{Conclusions} 

Polarization is a basic property of photons, just like frequency.  It is
often almost as densely coded with information!  Radio astronomers generally
make polarization observations automatically.  If optical astronomers did the
same, we might have some great discoveries.  Many years ago the polarimetry
optics absorbed a lot of photons, but now the total-flux spectrum accumulates
almost as fast with the polarimeter as without it.  Perhaps polarimetry
should be the default in the optical for some programs.  In particular,
I don't see why anyone would take total-flux spectra or images of distant AGN
or ULIRGs or Scuba sources when they
can get the polarization almost for free.

\begin{acknowledgments}
Thanks are due M.~Kishimoto, P.~Ogle, D.~Sanders, J.~Ulvestad, D.~Whysong,
and B.~Wills for comments on an earlier version of this paper.  I thank
Todd Hurt for use of his unpublished figure, and Makoto Kishimoto and
David Whysong for many significant contributions to the paper.
\end{acknowledgments}

%\appendix
%\section{Boundary conditions}


\begin{thebibliography}{} 

\bibitem[Agol \& Blaes (1996)]{Agol96}
{\sc Agol, E.\ \& Blaes, O.} 1996
{Polarization from magnetized accretion discs in active galactic nuclei.}
{\it MNRAS} {\bf 282}, 965.

\bibitem[Alloin {\it et al.}\/ (1985)]{Alloin85}
{\sc Alloin, D., Pelat, D., Phillips, M.\ \& Whittle, M.} 1985
{Recent spectral variations in the active nucleus of NGC 1566.}
 {\it ApJ.} {\bf 288}, 205.

\bibitem[Alloin, Barvainis, Gordon \& Antonucci (1992)]{Alloin92} 
{\sc Alloin, D., Barvainis, R., Gordon, M.\ A.\ \& Antonucci, R.\ R.\ J.} 1992 
{CO emission from radio quiet quasars - New detections support a thermal
                  origin for the FIR emission.}
{\it AAP} {\bf 265}, 429. 

\bibitem[Almaini, Lawrence \& Boyle (1999)]{Almaini99}
{\sc Almaini, O., Lawrence, A.\ \& Boyle, B.\ J.} 1999
{The AGN contribution to deep submillimetre surveys and the far-infrared
                  background.}
{\it MNRAS} {\bf 305}, L59. 

\bibitem[Antonucci (1982)]{Antonucci82} 
     {\sc Antonucci, R. R. J.} 1982(a)
{Optical polarization position angle versus radio source axis in radio galaxies.}
     {\it Natur.} {\bf 299}, 605.

\bibitem[Antonucci (1982)]{AntonucciPhD82} 
     {\sc Antonucci, R. R. J.} 1982(b)
{Optical flux and polarization spectra compared with radio maps of radio galaxies.}
     {\it Ph.D.\ thesis}, 1.

\bibitem[Antonucci (1983)]{Antonucci83} 
     {\sc Antonucci, R. R. J.} 1983
     {Optical polarization position angle versus radio structure axis in
 Seyfert galaxies.}
     {\it Natur.} {\bf 303}, 158.

\bibitem[Antonucci \& Ulvestad 1985]{Antonucci85}
   {\sc Antonucci, R.\ R.\ J.\ \& Ulvestad, J.\ S.} 1985
   {Extended radio emission and the nature of blazars.}
   {\it ApJ.} {\bf 294}, 158.

\bibitem[Antonucci 1988]{Antonucci88}
 {\sc Antonucci, R.} 1988,
{Polarization of active galactic nuclei and quasars.}
{\it Supermassive Black Holes}, 26.

\bibitem[Antonucci, Kinney \& Ford (1989)]{Antonucci89}
{\sc Antonucci, R.\ R.\ J., Kinney, A.\ L.\ \& Ford, H.\ C.} 1989
{The Lyman edge test of the quasar emission mechanism.}
{\it ApJ.} {\bf 342}, 64.

\bibitem[Antonucci \& Barvainis (1990)]{Antonucci90a}
{\sc Antonucci, R.\ \& Barvainis, R.} 1990
{Narrow-line radio galaxies as quasars in the sky plane.}
{\it ApJ. Lett.} {\bf 363}, L17.

\bibitem[Antonucci, Barvainis \& Alloin (1990)]{Antonucci90b} 
{\sc Antonucci, R., Barvainis, R.\ \& Alloin, D.} 1990
{The empirical difference between radio-loud and radio-quiet quasars.}
{\it ApJ.} {\bf 353}, 416.

\bibitem[Antonucci (1993)]{Antonucci93}
   {\sc Antonucci, R.} 1993
   {Unified models for active galactic nuclei and quasars.}
   {\it ARA\&A} {\bf 31}, 473. 

\bibitem[Antonucci, Kinney, \& Hurt(1993)]{1993ApJ...414..506A}
{\sc Antonucci, 
R., Kinney, A.\ L., \& Hurt, T.} 1993, {\it ApJ} {\bf 414}, 506.

\bibitem[Antonucci, Hurt \& Miller 1994]{Antonucci94}
 {\sc Antonucci, R., Hurt, T.\ \& Miller, J.} 1994
{HST ultraviolet spectropolarimetry of NGC 1068.}
 {\it ApJ.} {\bf 430}, 210.

\bibitem[Antonucci {\it et al.}\/ (1996)]{Antonucci96a}
{\sc Antonucci, R., Geller, R., Goodrich, R.\ W.\ \& Miller, J.\ S.} 1996,
 {The Spectropolarimetric Test of the Quasar Emission Mechanism.}
 {\it ApJ.} {\bf 472}, 502.

\bibitem[Antonucci, Hurt \& Agol 1996]{Antonucci96b}
 {\sc Antonucci, R., Hurt, T.\ \& Agol, E.} 1996
{Spectropolarimetric Test of the Relativistic Disk Model for the Broad H
 alpha Line of ARP 102B.}
 {\it ApJ. Lett.} {\bf 456}, L25.

\bibitem[Antonucci 1999]{Antonucci99}
 {\sc Antonucci, R.} 1999
 {Constraints on Disks Models of The Big Blue Bump from UV/Optical/IR
Observations.}
 {\it ASP Conf.\ Ser.\ 161: High Energy Processes in Accreting Black Holes},
 193.

\bibitem[Appenzeller {\it et al.}\/ 1998]{Appenzeller98}
 {\sc Appenzeller, I.\ {\it et al.\/}} 1998
{ORFEUS II Far-ultraviolet Observations of 3C 273: The Instrinsic Spectrum.}
 {\it ApJ. Lett.} {\bf 500}, L9.

\bibitem[Baker 1997]{Baker97}
   {\sc Baker, J.} 1997 {\it MNRAS} {\bf 286}, 23

\bibitem[Barthel 1987]{Barthel87}
   {\sc Barthel, P.\ D.} 1987 
{Feeling uncomfortable.}
   {\it Superluminal Radio Sources}, 148.

\bibitem[Barvainis 1992]{Barvainis92}
 {\sc Barvainis, R.} 1992
{Do accretion disks exist? IR through radio observations.}
 {\it Testing the AGN Paradigm}, 129.

\bibitem[Barvainis 1993]{Barvainis93}
 {\sc Barvainis, R.} 1993
{Free-free emission and the big blue bump in active galactic nuclei.}
 {\it ApJ.} {\bf 412}, 513.

\bibitem[Barvainis \& Lonsdale(1997)]{1997AJ....113..144B}
{\sc Barvainis, R.\ 
\& Lonsdale, C.} 1997, {\it AJ} {\bf 113}, 144.

\bibitem[Becker et al.(2000)]{2000ApJ...538...72B}
{\sc Becker, R.\ H., White, 
R.\ L., Gregg, M.\ D., Brotherton, M.\ S., Laurent-Muehleisen, S.\ A., \& 
Arav, N.} 2000, {\it ApJ} {\bf 538}, 72.

\bibitem[Berriman (1989)]{Berriman89}
{\sc Berriman, G.} 1989
{The origin of the optical polarizations of Seyfert 1 galaxies.}
{\it ApJ.} {\bf 345}, 713. 

\bibitem[Best {\it et al.}\/ 1997]{Best97}
   {\sc Best, P.\ N., Longair, M.\ S.\ \& Roettgering, J.\ H.\ A.} 1997
   {HST, radio and infrared observations of 28 3CR radio galaxies at
    redshift Z of about 1.}
   {\it MNRAS} {\bf 292}, 758.

\bibitem[Blaes \& Agol (1996)]{Blaes96}
{\sc Blaes, O.\ \& Agol, E.} 1996
{Polarization near the Lyman Edge in Accretion Disk Atmophere Models of Quasars.}
{\it ApJ. Lett.} {\bf 469}, L41.

%\bibitem[Bregman {\it et al.\/} (1988)]{Bregman88}
%   {\sc Bregman, J.\ N.\ {\it et al.\/}} 1988
%{Multifrequency observation of the optically violent variable quasar 3C 446.}
%   {\it ApJ.} {\bf 331}, 746. 

\bibitem[Brindle {\it et al.}\/ (1990)]{Brindle90}
{\sc Brindle, C., Hough, J.\ H., Bailey, J.\ A., Axon, D.\ J., Ward, M.\ J.,
Sparks, W.\ B.\ \& McLean, I.\ S.} 1990
{An Optical and Near Infrared Polarization Survey of Seyfert and Broadline
Radio Galaxies - Part Two - the Wavelength Dependence of Polarization.}
{\it MNRAS} {\bf 244}, 604.

\bibitem[Brotherton et al.(1997)]{1997ApJ...487L.113B}
{\sc Brotherton, M.\ S., 
Tran, H.\ D., van Breugel, W., Dey, A., \& Antonucci, R.} 1997,
 {\it ApJ. Lett.} {\bf 487}, L113.

\bibitem[Browne {\it et al.\/} 1982]{Browne82}
   {\sc Browne, I.\ W.\ A., Clark, R.\ R., Moore, P.\ K., Muxlow, T.\ W.\ B.,
        Wilkinson, P.\ N., Cohen, M.\ H.\ \& Porcas, R.\ W.} 1982
   {MERLIN observations of superluminal radio sources.}
   {\it Natur.} {\bf 299}, 788.

\bibitem[Capetti {\it et al.}\/ (1995a)]{Capetti95a}
{\sc Capetti, A., Axon, D.\ J., Macchetto, F., Sparks, W.\ B.\ \&
 Boksenberg, A.} 1995
{HST Imaging Polarimetry of NGC 1068.}
{\it ApJ.} {\bf 446}, 155.

\bibitem[Capetti {\it et al.}\/ (1995b)]{Capetti95b}
{\sc Capetti, A., Macchetto, F., Axon, D.\ J., Sparks, W.\ B.\ \&
 Boksenberg, A.} 1995
{Hubble Space Telescope Imaging Polarimetry of the Inner Nuclear Region of
                  NGC 1068.}
{\it ApJ. Lett.} {\bf 452}, L87. 

\bibitem[Capetti {\it et al.\/} 2000]{Capetti00}
   {\sc Capetti, A.\ {\it et al.\/}} 2000
   {Hubble Space Telescope Infrared Imaging Polarimetry of Centaurus A:
    Implications for the Unified Scheme and the Existence of a Misdirected
    BL Lacertae Nucleus.}
   {\it ApJ.} {\bf 544}, 269.

\bibitem[Chen \& Halpern (1990)]{Chen90}
 {\sc Chen, K.\ \& Halpern, J.\ P.} 1990
 {Spectropolarimetric test of the relativistic disk model for the broad
emission lines of active galactic nuclei.}
 {\it ApJ. Lett.} {\bf 354}, L1.

\bibitem[Chen, Halpern \& Titarchuk 1997]{Chen97}
 {\sc Chen, K., Halpern, J.\ P.\ \& Titarchuk, L.\ G.} 1997
 {Polarization of Line Emission from an Accretion Disk and Application to
ARP 102B.}
 {\it ApJ.} {\bf 483}, 194.

\bibitem[Chiaberge {\it et al.}\/ 1999]{Chiaberge99}
   {\sc Chiaberge, M., Capetti, A.\ \& Celotti, A.} 1999
   {The HST view of FR I radio galaxies: evidence for non-thermal nuclear
    sources.}
   {\it AAP} {\bf 349}, 77.

\bibitem[Chiaberge {\it et al.}\/ (2000)]{Chiaberge00}
   {\sc Chiaberge, M., Capetti, A.\ \& Celotti, A.} 2000
   {The HST view of the FR I / FR II dichotomy.}
   {\it AAP} {\bf 355}, 873.

\bibitem[Cimatti {\it et al.}\/ (1996)]{Cimatti96}
{\sc Cimatti, A., Dey, A., van Breugel, W., Antonucci, R., and Spinrod, H.}
1996
{\it ApJ} {\bf 465}, 145.

\bibitem[Cohen {\it et al.}\/ (1999)]{Cohen99}
{\sc Cohen, M.\ H., Ogle, P.\ M., Tran, H.\ D., Goodrich, R.\ W.\ \&
 Miller, J.\ S.} 1999
{Polarimetry and Unification of Low-Redshift Radio Galaxies.}
{\it A.J.} {\bf 118}, 1963. 

\bibitem[Collin-Souffrin {\it et al.}\/ (1988)]{Collin88}
{\sc Collin-Souffrin, S., Dyson, J.\ E.,
 McDowell, J.\ C.\ \& Perry, J.\ J.} 1988
{The environment of active galactic nuclei. I - A two-component broad
                  emission line model.}
{\it MNRAS} {\bf 232}, 539.

\bibitem[Corbett {\it et al.}\/ 1998]{Corbett98}
 {\sc Corbett, E.\ A., Robinson, A., Axon, D.\ J., Young, S.\ \&
 Hough, J.\ H.} 1998
 {The profiles of polarized broad Halpha lines in radio galaxies.}
 {\it MNRAS} {\bf 296}, 721.

\bibitem[Cutri {\it et al.}\/ (1985)]{Cutri85} 
{\sc Cutri, R.\ M., Wisniewski, W.\ Z., Rieke, G.\ H.\ \& Lebofsky, M.\ J.} 
1985
{Variability and the nature of QSO optical-infrared continua.}
 {\it ApJ.} {\bf 296}, 423.

\bibitem[Czerny \& Pojmanski 1990]{Czerny90}
 {\sc Czerny, B.\ \& Pojmanski, G.} 1990
{Lyman edges in AGN accretion discs.}
 {\it MNRAS} {\bf 245}, 1P.

\bibitem[Czerny \& Zbyszewska 1991]{Czerny91}
 {\sc Czerny, B.\ \& Zbyszewska, M.} 1991
{Comptonization of the Lyman edge in active galactic nuclei.}
 {\it MNRAS} {\bf 249}, 634.

\bibitem[Czerny, Jaroszynski \& Czerny (1994)]{Czerny94}
 {\sc Czerny, B., Jaroszynski, M.\ \& Czerny, M.} 1994
{Constraints on the Size of the Emitting Region in an Active Galactic Nucleus.}
 {\it MNRAS} {\bf 268}, 135.

\bibitem[Dey {\it et al.}\/ 1996]{Dey96} 
   {\sc Dey, A., Cimatti, A., van Breugel, W., Antonucci, R. and Spinrod,
H.} 1996
{\it ApJ} {\bf 465}, 157.

\bibitem[Dey {\it et al.}\/ 1997]{Dey97} 
   {\sc Dey, A., van Breugel, W., Vacca, W.\ D.\ \& Antonucci, R.} 1997
   {Triggered Star Formation in a Massive Galaxy at Z = 3.8: 4C 41.17.}
   {\it ApJ.} {\bf 490}, 698.

\bibitem[Done {\it et al.\/} 1990]{Done90}
 {\sc Done, C., Ward, M.\ J., 
Fabian, A.\ C., Kunieda, H., Tsuruta, S., Lawrence, A., Smith, M.\ G.\ \& 
Wamsteker, W.} 1990
{Simultaneous multifrequency observations of the Seyfert 1 galaxy NGC
4051 -
Constant optical-infrared emission observed during large-amplitude
X-ray variability.}
 {\it MNRAS} {\bf 243}, 713.

\bibitem[Ekers \& Simkin 1983]{Ekers83}
   {\sc Ekers, R.\ D.\ \& Simkin, S.\ M.} 1983
   {Radio structure and optical kinematics of the cD galaxy Hydra A /3C 218/.}
   {\it ApJ.} {\bf 265}, 85.

\bibitem[Elvius (1978)]{Elvius78}
{\sc Elvius, A.} 1978
{Polarization of light in the Seyfert galaxy NGC 1068.}
{\it AAP} {\bf 65}, 233. 

\bibitem[Eracleus and Halpern (2001)]{Eracleus01}
{\sc Eracleus, M.\ and Halpern, J.} 2001
{A certified LINER with broad variable emission lines.}
{\it Ap. J.} (in press); also astro-ph/0101050.

\bibitem[Evans {\it et al.}\/ (2001)]{Evans01}
{\sc Evans, A.\ S., Frayer, Surace, J.\ A., Sanders,  D.\ B.} 2001
{Molecular Gas in Infrared-Excess, Optically-Selected QSOs and the
Connection with Infrared Luminous Galaxies.}
{\it AJ}, in press - also astro-ph 0101308.

\bibitem[Falcke, Sherwood, \& Patnaik(1996)]{1996ApJ...471..106F}
{\sc Falcke, 
H., Sherwood, W., \& Patnaik, A.\ R.} 1996, {\it ApJ} {\bf 471}, 106.

\bibitem[Francis 1996]{Francis96}
{\sc Francis, P.\ J.} 1996
{The continuum slopes of optically selected QSOs.}
{\it Publications of the Astronomical Society of Australia} {\bf 13}, 212.

\bibitem[Geller {\it et al.}\/ (2000)]{Geller00}
{\sc Geller, R.\ M., Sault, R.\ J., Antonucci, R., Killeen, N.\ E.\ B.,
Ekers, R., Desai, K.\ \& Whysong, D.} 2000
{Cosmological Halos: A Search for the Ionized Intergalactic Medium.}
{\it ApJ.} {\bf 539}, 73.

\bibitem[Genzel {\it et al.}\/ (1998)]{Genzel98}
{\sc Genzel, R.} {\it et al.}\/ 1998
{What Powers Ultraluminous IRAS Galaxies?}
{\it ApJ.} {\bf 498}, 579. 

\bibitem[Giveon {\it et al.\/} 1999]{Giveon99}
 {\sc Giveon, U., Maoz, D., Kaspi, S., Netzer, H.\ \& Smith, P.\ S.} 1999
{Long-term optical variability properties of the Palomar-Green quasars.}
 {\it MNRAS} {\bf 306}, 637.

\bibitem[Gnedin \& Silantev (1978)]{Gnedin78}
{\sc Gnedin, I.\ N.\ \& Silantev, N.\ A.} 1978
{Polarization effects in the emission of a disk of accreting matter.}
{\it Soviet Astronomy} {\bf 22}, 325.

\bibitem[Gonz{\'a}lez Delgado {\it et al.}\/ (1998)]{Gonzalez98} 
{\sc Gonz{\'a}lez Delgado, R.\ M., Heckman, T., Leitherer, C., Meurer, G., 
Krolik, J., Wilson, A.\ S., Kinney, A.\ \& Koratkar, A.} 1998
{Ultraviolet-Optical Observations of the Seyfert 2 Galaxies NGC 7130, NGC
5135, and IC 3639: Implications for the Starburst-Active Galactic Nucleus
                  Connection.}
{\it ApJ.} {\bf 505}, 174.

\bibitem[Goodrich (1989)]{Goodrich89} 
     {\sc Goodrich, Robert W.} 1989
     {Spectropolarimetry of `narrow-line' Seyfert 1 galaxies.}
     {\it ApJ.} {\bf 342}, 224. 

\bibitem[Goodrich \& Miller (1994)]{Goodrich94}
{\sc Goodrich, R.\ W.\ \& Miller, J.\ S.} 1994
{Spectropolarimetry of high-polarization Seyfert 1 galaxies: Geometry and
                  kinematics of the scattering regions.}
{\it ApJ.} {\bf 434}, 82.

\bibitem[Goodrich {\it et al.}\/ (1996)]{Goodrich96}
{\sc Goodrich, R.\ W., Miller, J.\ S., Martel, A., Cohen, M.\ H.,
Tran, H.\ D., Ogle, P.\ M.\ \& Vermeulen, R.\ C.} 1996
{FSC 10214+4724: A Gravitationally Lensed, Hidden QSO.}
{\it ApJ. Lett.} {\bf 456}, L9.

\bibitem[Goodrich(1997)]{1997ApJ...474..606G}
{\sc Goodrich, R.\ W.} 1997,
 {\it ApJ.} {\bf 474}, 606.

\bibitem[Gopal-Krishna {\it et al.}\/ (1996)]{Gopal-Krishna96}
   {\sc Gopal-Krishna, Kulkarni, V.\ K.\ \& Wiita, P.\ J.} 1996
   {The Linear Sizes of Quasars and Radio Galaxies in the Unified Scheme.}
   {\it ApJ. Lett.} {\bf 463}, L1.

\bibitem[Gopal-Krishna \& Biermann 1998]{Gopal-Krishna98} 
{\sc Gopal-Krishna \& Biermann, P.\ L.} 1998
{Are ultra-luminous infrared galaxies the dominant extragalactic population at
                  high luminosities?}
{\it AAP} {\bf 330}, L37. 

\bibitem[Gregg et al.(2000)]{2000ApJ...544..142G}
{\sc Gregg, M.\ D., Becker, 
R.\ H., Brotherton, M.\ S., Laurent-Muehleisen, S.\ A., Lacy, M., \& White, 
R.\ L.} 2000, {\it ApJ} {\bf 544}, 142.

\bibitem[Halpern, Eracleous, Filippenko, \& 
Chen(1996)]{1996ApJ...464..704H}
{\sc Halpern, J.\ P., Eracleous, M., 
Filippenko, A.\ V., \& Chen, K.} 1996, {\it ApJ} {\bf 464}, 704.

\bibitem[Heckman {\it et al.}\/ (1995)]{Heckman95}
{\sc Heckman, T.} {\it et al.}\/ 1995
{The Nature of the Ultraviolet Continuum in Type 2 Seyfert Galaxies.}
{\it ApJ.} {\bf 452}, 549.

\bibitem[Heisler, Lumsden \& Bailey (1997)]{Heisler97}
{\sc Heisler, C.\ A., Lumsden, S.\ L.\ \& Bailey, J.\ A.} 1997
{Visibility of scattered broad-line emission in Seyfert 2 galaxies.}
{\it Natur.} {\bf 385}, 700.

\bibitem[Hines {\it et al.}\/ (1999)]{Hines00}
{\sc Hines, D.\ C., Schmidt, G.\ D., Wills, B.\ J., Smith, P.\ S., and
Sowindki, L.\ G.} 1999
{\it ApJ} {\bf 512}, 145.

\bibitem[Hurt {\it et al.\/} 1999]{Hurt99}
   {\sc Hurt, T., Antonucci, R., Cohen, R., Kinney, A.\ \& Krolik, J.} 1999
   {Ultraviolet Imaging Polarimetry of Narrow-Line Radio Galaxies.}
   {\it ApJ.} {\bf 514}, 579.

\bibitem[Hutsem{\'e}kers \& Lamy(2000)]{2000A&A...358..835H} 
{\sc Hutsem{\'e}kers, D.\ \& Lamy, H.} 2000, {\it AAP} {\bf 358}, 

\bibitem[Impey \& Neugebauer (1988)]{Impey88}
{\sc Impey, C.\ D.\ \& Neugebauer, G.} 1988
{Energy distributions of blazars.}
{\it A.J.} {\bf 95}, 307. 

\bibitem[Jaffe {\it et al.\/} (1996)]{Jaffe96}
   {\sc Jaffe, W., Ford, H., Ferrarese, L., van den Bosch, F.\ \& O'Connell,
        R.\ W.} 1996
   {The Nuclear Disk of NGC 4261: Hubble Space Telescope Images and
    Ground-based Spectra.}
   {\it ApJ.} {\bf 460}, 214. 

\bibitem[Kay (1994)]{Kay94}
{\sc Kay, L.\ E.} 1994
{Blue spectropolarimetry of Seyfert 2 galaxies. 1: Analysis and basic results.}
{\it ApJ.} {\bf 430}, 196.

\bibitem[Kishimoto (1999)]{Kishimoto99}
{\sc Kishimoto, M.} 1999
{The Location of the Nucleus of NGC 1068 and the Three-dimensional
Structure of Its Nuclear Region.}
{\it ApJ.} {\bf 518}, 676.

\bibitem[Kishimoto {\it et al.}\/ (2000)]{Kishimoto00}
   {\sc Kishimoto, M., Antonucci. R., Cimatti, A., Hurt, T., Dey, A.,
        van Breugel, W., Spinrad, H.} 2000
   {UV Spectropolarimetry of Narrow-line Radio Galaxies.}

\bibitem[Knapp \& Patten (1991)]{Knapp91}
{\sc Knapp, G.\ R.\ \& Patten, B.\ M.} 1991
{Millimeter and submillimeter observations of nearby radio
galaxies.}
{\it A.J.} {\bf 101}, 1609.

\bibitem[Kollgaard {\it et al.}\/ 1992]{Kollgaard92}
   {\sc Kollgaard, R. I., Wardle, J. F. C., Roberts, D. H.,
        Gabuzda, D. C.} 1992
   {Radio constraints on the nature of BL Lacertae objects and their
    parent population.}
   {\it Astronomical Journal} {\bf vol.~104}, p.~1687.

\bibitem[Kolykhalov \& Sunyaev (1984)]{Kolykhalov84}
 {\sc Kolykhalov, P.\ I.\ \& Sunyaev, R.\ A.} 1984
{Radiation of accretion disks in quasars and galactic nuclei.}
 {\it Advances in Space Research} {\bf 3}, 249.

\bibitem[Koratkar {\it et al.\/} 1995]{Koratkar95}
 {\sc Koratkar, A., Antonucci, R.\ R.\ J., Goodrich, R.\ W., Bushouse, H.\ \&
 Kinney, A.\ L.} 1995
 {Quasar Lyman Edge Regions in Polarized Light.}
 {\it ApJ.} {\bf 450}, 501.

\bibitem[Koratkar {\it et al.}\/ (1998)]{Koratkar98}
{\sc Koratkar, A., Antonucci, R., Goodrich, R.\ \& Storrs, A.} 1998
{Below the Lyman Edge: Ultraviolet Polarimetry of Quasars.}
 {\it ApJ.} {\bf 503}, 599.

\bibitem[Koratkar \& Blaes (1999)]{Koratkar99}
{\sc Koratkar, A.\ \& Blaes, O.} 1999
{The Ultraviolet and Optical Continuum Emission in Active Galactic Nuclei: The
                  Status of Accretion Disks.}
{\it PASP} {\bf 111}, 1.

\bibitem[Krolik \& Begelman (1986)]{Krolik86}
{\sc Krolik, J.\ H.\ \& Begelman, M.\ C.} 1986
{An X-ray heated wind in NGC 1068.}
{\it ApJ. Lett.} {\bf 308}, L55. 

\bibitem[Krolik \& Begelman (1988)]{Krolik88}
{\sc Krolik, J.\ H.\ \& Begelman, M.\ C.} 1988
{Molecular tori in Seyfert galaxies - Feeding the monster and hiding it.}
{\it ApJ.} {\bf 329}, 702. 

\bibitem[Krolik {\it et al.\/} 1991]{Krolik91}
 {\sc Krolik, J.\ H., Horne, K., Kallman, T.\ R., Malkan, M.\ A.,
 Edelson, R.\ A.\ \& Kriss, G.\ A.} 1991
 {Ultraviolet variability of NGC 5548 - Dynamics of the continuum
production region and geometry of the broad-line region.}
 {\it ApJ.} {\bf 371}, 541.

\bibitem[Krolik \& Voit(1998)]{1998ApJ...497L...5K}
{\sc Krolik, J.\ H.\ \& 
Voit, G.\ M.} 1998, {\it ApJ. Lett.} {\bf 497}, L5.

\bibitem[Laing 1988]{Laing88}
   {\sc Laing, R.\ A.} 1988
   {The sidedness of jets and depolarization in powerful extragalactic radio
    sources.}
   {\it Natur.} {\bf 331}, 149.

\bibitem[Lamy \& Hutsem{\'e}kers(2000)]{2000A&A...356L...9L}
{\sc Lamy, H.\ \& 
Hutsem{\'e}kers, D.} 2000, {\it AAP} {\bf 356}, L9.

\bibitem[Landau {\it et al.\/} (1986)]{Landau86}
   {\sc Landau, R.\ {\it et al.\/}} 1986
   {Active extragalactic sources --- Nearly simultaneous observations from 20
                  centimeters to 1400\AA.}
   {\it ApJ.} {\bf 308}, 78.

\bibitem[Laor (1990)]{Laor90}
 {\sc Laor, A.} 1990
{AGN Accretion Discs --- Part Three --- Comparison with the Observations.}
 {\it MNRAS} {\bf 246}, 369.

\bibitem[Laor, Jannuzi, Green \& Boroson (1997)]{Laor97}
{\sc Laor, A., Jannuzi, B.\ T., Green, R.\ F.\ \& Boroson, T.\ A.} 1997
{The Ultraviolet Properties of the Narrow-Line Quasar I ZW 1.}
{\it ApJ.} {\bf 489}, 656.

\bibitem[Lara {\it et al.\/} 1999]{Lara99}
   {\sc Lara, L., M{\'a}rquez, I., Cotton, W.\ D., Feretti, L.,
        Giovannini, G., Marcaide, J.\ M.\ \& Venturi, T.} 1999
   {The broad-line radio galaxy J2114+820.}
   {\it New Astronomy Review} {\bf 43}, 643.

\bibitem[Leighly 2000]{Leighly00}
{\sc Leighly, K.\ M.} 2000
{STIS Ultraviolet Spectral Evidence for
Outflows in Extreme Narrow-Line Seyfert 1 Galaxies.}
{\it astro-ph/0012173}

\bibitem[Lutz, Veilleux \& Genzel (1999)]{Lutz99}
{\sc Lutz, D., Veilleux, S.\ \& Genzel, R.} 1999
{Mid-Infrared and Optical Spectroscopy of Ultraluminous Infrared Galaxies: A
                  Comparison.}
{\it ApJ. Lett.}, 517, L13.

\bibitem[Madejski {\it et al.}\/ (2000]{Madejski00}
{\sc Madejski, G.\ M., {\.Z}ycki, P., Done, C., Valinia, A., Blanco, P.,
Rothschild, R.\ \& Turek, B.} 2000
{Structure of the Circumnuclear Region of Seyfert 2 Galaxies Revealed by Rossi
                  X-Ray Timing Explorer Hard X-Ray Observations of NGC 4945.}
{\it ApJ. Lett.} {\bf 535}, 87.

\bibitem[Maiolino {\it et al.}\/ (1998)]{Maiolino98}
{\sc Maiolino, R., Salvati, M., Bassani, L., Dadina, M., della Ceca, R.,
 Matt, G., Risaliti, G.\ \& Zamorani, G.} 1998.
{Heavy obscuration in X-ray weak AGNs.}
{\it AAP} {\bf 338}, 781. 

\bibitem[Malkan (1983)]{Malken83}
{\sc Malkan, M.\ A.} 1983
{The ultraviolet excess of luminous quasars. II - Evidence for massive
accretion disks.}
{\it ApJ.} {\bf 268}, 582.

\bibitem[Marconi {\it et al.\/} 2000]{Marconi00}
   {\sc Marconi, A., Schreier, E.\ J., Koekemoer, A., Capetti, A., Axon, D.,
        Macchetto, D.\ \& Caon, N.} 2000
   {Unveiling the Active Nucleus of Centaurus A.}
   {\it ApJ.} {\bf 528}, 276.

\bibitem[Martel (1996)]{Martel96}
{\sc Martel, A.\ R.} 1996
{Spectropolarimetry of high-polarization Seyfert 1 galaxies.}
{\it Ph.D.\ Thesis}, 80. 

\bibitem[Martel (1998)]{Martel98}
{\sc Martel, A.} 1998
{New H$\alpha$ Spectropolarimetry of NGC 4151: The Broad-Line Region-Host
                  Connection.}
{\it ApJ.} {\bf 508}, 657.

\bibitem[Martin {\it et al.}\/ (1983)]{Martin83} 
     {\sc Martin, P. G., Thompson, I. B., Maza, J., \& Angel, J. R. P.} 1983 
     {The polarization of Seyfert galaxies.} 
     {\it ApJ.} {\bf 266}, 470. 

\bibitem[Mathewson \& Ford (1970)]{Mathewson70}
{\sc Mathewson, D.\ S.\ \& Ford, V.\ L.} 1970
{Polarization observations of 1800 stars.}
{\it MEMRAS} {\bf 74}, 139.

\bibitem[Matt, Fabian \& Ross 1993]{Matt93}
 {\sc Matt, G., Fabian, A.\ C.\ \& Ross, R.\ R.} 1993
{X-Ray Photoionized Accretion Discs - Ultraviolet and X-Ray Continuum Spectra
 and Polarization.}
 {\it MNRAS} {\bf 264}, 839.

\bibitem[McLean {\it et al.}\/ 1983]{McLean83} 
     {\sc McLean, I. S., Aspin, C., Heathcote, S. R. \& McCaughrean, M. J.}
     1983
     {Is the polarization of NGC1068 evidence for a non-thermal source?} 
     {\it Natur.} {\bf 304}, 609. 

\bibitem[Miller \& Antonucci 1983]{Miller83} 
     {\sc Miller, J. S. \& Antonucci, R. R. J.} 1983 
     {Evidence for a highly polarized continuum in the nucleus of NGC 1068.} 
     {\it ApJ. Lett.} {\bf 271}, 7. 

\bibitem[Miller \& Goodrich (1990)]{Miller90}
{\sc Miller, J.\ S.\ \& Goodrich, R.\ W.} 1990
{Spectropolarimetry of high-polarization Seyfert 2 galaxies and unified Seyfert
 theories}
{\it ApJ.} {\bf 355}, 456.

\bibitem[Miller, Goodrich \& Mathews (1991)]{Miller91}
{\sc Miller, J.\ S., Goodrich, R.\ W.\ \& Mathews, W.\ G.} 1991
{Multidirectional views of the active nucleus of NGC 1068.}
{\it ApJ.} {\bf 378}, 47.

\bibitem[Mineo {\it et al.}\/ (2000)]{Mineo00}
{\sc Mineo, T.} {\it et al.}\/ 2000
{BeppoSAX broad-band observations of low-redshift quasars: spectral
                  curvature and iron Kalpha lines.}
{\it AAP} {\bf 359}, 471.

\bibitem[Miyaji, Wilson \& Perez-Fournon (1992)]{Miyaji92} 
{\sc Miyaji, T., Wilson, A.\ S.\ \& Perez-Fournon, I.} 1992
{The radio source and bipolar nebulosity in the Seyfert galaxy NGC 3516.}
{\it ApJ.} {\bf 385}, 137. 

\bibitem[Mushotzky \& Wandel 1989]{Mushotzky89}
 {\sc Mushotzky, R.\ F.\ \& Wandel, A.} 1989
{On the ratio of the infrared-to-ultraviolet continuum to the X-rays in
 quasars and active galaxies.}
 {\it ApJ.} {\bf 339}, 674.

\bibitem[Nandra {\it et al.\/} (1998)]{Nandra98}
 {\sc Nandra, K., Clavel, J., 
Edelson, R.\ A., George, I.\ M., Malkan, M.\ A., Mushotzky, R.\ F., 
Peterson, B.\ M.\ \& Turner, T.\ J.} 1998
{New Constraints on the Continuum Emission Mechanism of Active Galactic Nuclei:
Intensive Monitoring of NGC 7469 in the X-Ray and Ultraviolet.}
 {\it ApJ.} {\bf 505}, 594.

\bibitem[Nandra {\it et al.\/} (2000)]{Nandra00}
 {\sc Nandra, K., Le, T., 
George, I.\ M., Edelson, R.\ A., Mushotzky, R.\ F., Peterson, B.\ M.\ \& 
Turner, T.\ J.} 2000
{Origin of the X-Ray and Ultraviolet Emission in NGC 7469.}
 {\it ApJ.} {\bf 544}, 734.

\bibitem[Neugebauer {\it et al.\/} 1987]{Neugebauer87}
 {\sc Neugebauer, G., Green, R.\ F., Matthews, K., Schmidt, M.,
 Soifer, B.\ T.\ \& Bennett, J.} 1987
{Continuum energy distributions of quasars in the Palomar-Green Survey.}
 {\it ApJ. Supp.} {\bf 63}, 615.

\bibitem[Ogle et al.(1999)]{1999ApJS..125....1O}
{\sc Ogle, P.\ M., Cohen, M.\ 
H., Miller, J.\ S., Tran, H.\ D., Goodrich, R.\ W., \& Martel, A.\ R.} 
1999, {\it ApJ. Supp} {\bf 125}, 1.

\bibitem[Owen \& Ledlow (1994)]{Owen94}
   {\sc Owen, F.\ N.\ \& Ledlow, M.\ J.} 1994
   {The First Stromlo Symposium: The Physics of Active Galaxies.}
   ASP Conference Series, Vol.\ 54, 1994, G.V.\ Bicknell, M.A.\ Dopita,
   and P.J.\ Quinn, Eds., p.~319.

\bibitem[Owen, Eilek \& Kassim (2000)]{Owen00}
{\sc Owen, F.\ N., Eilek, J.\ A.\ \& Kassim, N.\ E.} 2000
{M87 at 90 Centimeters: A Different Picture.}
{\it ApJ.} {\bf 543}, 611.

\bibitem[Pier \& Krolik (1993)]{Pier93}
{\sc Pier, E.\ A.\ \& Krolik, J.\ H.} 1993
{Infrared Spectra of Obscuring Dust Tori around Active Galactic Nuclei. II.
                  Comparison with Observations.}
{\it ApJ.} {\bf 418}, 673.

\bibitem[Rauch \& Blandford (1991)]{Rauch91}
 {\sc Rauch, K.\ P.\ \& Blandford, R.\ D.} 1991
{Microlensing and the structure of active galactic nucleus accretion
disks.}
 {\it ApJ. Lett.} {\bf 381}, L39.

\bibitem[Reimers {\it et al.\/} 1989]{Reimers89}
 {\sc Reimers, D., Clavel, J., Groote, D., Engels, D., Hagen, H.\ J.,
Naylor, T., Wamsteker, W.\ \& Hopp, U.} 1989
{The luminous quasar HS1700+6416 and the shape of the `big bump' below 500 A.}
 {\it AAP} {\bf 218}, 71.

\bibitem[Reynolds {\it et al.\/} 1996]{Reynolds96}
   {\sc Reynolds, C.\ S., di Matteo, T., Fabian, A.\ C., Hwang, U.\ \&
        Canizares, C.\ R.} 1996
   {The `quiescent' black hole in M87.}
   {\it MNRAS} {\bf 283}, L111.

\bibitem[Rodriguez-Pascual, Mas-Hesse \& Santos-Lleo (1997)]{Rodriguez97}
{\sc Rodriguez-Pascual, P.\ M., Mas-Hesse, J.\ M.\ \& Santos-Lleo, M.} 1997
{The broad line region of narrow-line Seyfert 1 galaxies.}
{\it AAP} {\bf 327}, 72. 

\bibitem[Rydbeck {\it et al.\/} 1993]{Rydbeck93}
   {\sc Rydbeck, G., Wiklind, T., Cameron, M., Wild, W., Eckart, A.,
        Genzel, R.\ \& Rothermel, H.} 1993
   {High resolution (C-12)O(2-1) observations of the molecular gas in
    Centaurus A.}
   {\it AAP} {\bf 270}, L13.

\bibitem[Salvati \& Maiolino (2000)]{Salvati00}
{\sc Salvati, M.\ \& Maiolino, R.} 2000
{Where are the Type 2 AGNs?}
{\it Large Scale Structure in the X-ray Universe}, 
Proceedings of the 20--22 September 1999 Workshop, Santorini, Greece, eds.\ 
Plionis, M.\ \& Georgantopoulos, I., Atlantisciences, Paris, France, p.~277 

\bibitem[Sambruna {\it et al.\/} 2000]{Sambruna00}
   {\sc Sambruna, R.\ M., Chartas, G., Eracleous, M., Mushotzky, R.\ F.\ \&
        Nousek, J.\ A.} 2000
   {Chandra Uncovers a Hidden Low-Luminosity Active Galactic Nucleus
    in the Radio Galaxy Hydra A (3C 218).}
   {\it ApJ Lett.} {\bf 532}, L91.

\bibitem[Sanders {\it et al.}\/ (1989)]{Sanders89}
{\sc Sanders, D.\ B., Phinney, E.\ S., Neugebauer, G., Soifer, B.\ T.\ \&
Matthews, K.} 1989
{Continuum Energy Distributions of Quasars: Shapes and Origins.}
{\it ApJ.} {\bf 347}, 29.

\bibitem[Sarazin {\it et al.\/} 1999]{Sarazin99}
   {\sc Sarazin, C.\ L., Koekemoer, A.\ M., Baum, S.\ A., O'Dea, C.\ P.,
        Owen, F.\ N.\ \& Wise, M.\ W.} 1999
   {X-Ray Properties of B2 1028+313: A Quasar at the Center of the Abell
    Cluster A1030.}
   {\it ApJ.} {\bf 510}, 90.

\bibitem[Scargle, Caroff, \& Noerdlinger(1970)]{1970ApJ...161L.115S} 
{\sc Scargle, J.\ D., Caroff, L.\ J., \& Noerdlinger, P.\ D.} 1970,
 {\it ApJ. Lett.} {\bf 161}, L115.

\bibitem[Schmidt \& Miller (1980)]{Schmidt80}
{\sc Schmidt, G.\ D.\ \& Miller, J.\ S.} 1980
{The spectrum and polarization of the nucleus of NGC 4151.}
{\it ApJ.}, {\bf 240}, 759.

\bibitem[Schmidt \& Hines(1999)]{1999ApJ...512..125S}
{\sc Schmidt, G.\ D.\ \& 
Hines, D.\ C.} 1999, {\it ApJ} {\bf 512}, 125.

\bibitem[Schmidt \& Smith 2000]{Schmidt00}
{\sc Schmidt, Gary D.\ \& Smith, Paul S.} 2000
{Evidence for Polarized Synchrotron Components in Radio-Optical Aligned
Quasars.}
 {\it ApJ.} in press, or astro-ph0008168.

\bibitem[Siemiginowska \& Dobrzycki 1990]{Siemiginowska90}
{\sc Siemiginowska, A.\ \& Dobrzycki, A.} 1990
{Accretion disk in the high-redshift quasar HS 1700 + 6416.}
 {\it AAP} {\bf 231}, L1.

\bibitem[Sincell \& Krolik 1997]{Sincell97}
 {\sc Sincell, M.\ W.\ \& Krolik, J.\ H.} 1997,
 {The Vertical Structure and Ultraviolet Spectrum of
X-Ray--irradiated Accretion Disks in Active Galactic Nuclei.}
 {\it ApJ.} {\bf 476}, 605.

\bibitem[Singal (1993)]{Singal93}
   {\sc Singal, A.\ K.} 1993
   {Evidence against the unified scheme for powerful radio galaxies and
    quasars.}
   {\it MNRAS} {\bf 262}, L27.

\bibitem[Singh \& Westergaard (1992)]{Singh92}
{\sc Singh, K.\ P.\ \& Westergaard, N.\ J.} 1992
{Radio structure of the Seyfert galaxy Markarian 509.}
{\it AAP} {\bf 264}, 489. 

\bibitem[Stockman, Angel, \& Miley 1979]{Stockman79} 
     {\sc Stockman, H. S., Angel, J. R. P., \& Miley, G. K.} 1979
     {Alignment of the optical polarization with the radio structure of
QSOs.}
     {\it ApJ. Lett.} {\bf 227}, 55. 

\bibitem[Stockman, Hier, \& Angel(1981)]{1981ApJ...243..404S}
{\sc Stockman, H.\ 
S., Hier, R.\ G., \& Angel, J.\ R.\ P.} 1981, {\it ApJ} {\bf 243}, 404.

\bibitem[Surace, Sanders \& Evans (2000)]{Surace00}
{\sc Surace, J.\ A., Sanders, D.\ B.\ \& Evans, A.\ S.} 2000
{High-Resolution Optical/Near-Infrared Imaging of Cool Ultraluminous Infrared
                  Galaxies.}
{\it ApJ.} {\bf 529}, 170. 

\bibitem[Thompson \& Martin (1988)]{Thompson88}
{\sc Thompson, I.\ B.\ \& Martin, P.\ G.} 1988
{Optical polarization of Seyfert galaxies.}
{\it ApJ.} {\bf 330}, 121. 

\bibitem[Tran, Miller \& Kay (1992)]{Tran1992}
{\sc Tran, H.\ D., Miller, J.\ S.\ \& Kay, L.\ E.} 1992
{Detection of obscured broad-line regions in four Seyfert 2 galaxies.}
{\it ApJ.} {\bf 397}, 452.

\bibitem[Tran (1995a)]{Tran95a}
{\sc Tran, H.\ D.} 1995(a)
{The Nature of Seyfert 2 Galaxies with Obscured Broad-Line Regions. II.
                  Individual Objects.}
{\it ApJ.} {\bf 440}, 578.

\bibitem[Tran (1995b)]{Tran95b}
{\sc Tran, H.\ D.} 1995(b)
{The Nature of Seyfert 2 Galaxies with Obscured Broad-Line Regions. III.
                  Interpretation.}
{\it ApJ.} {\bf 440}, 597.

\bibitem[Tran, Cohen \& Goodrich (1995)]{Tran95}
{\sc Tran, H.\ D., Cohen, M.\ H.\ \& Goodrich, R.\ W.} 1995
{Keck Spectropolarimetry of the Radio Galaxy 3C 234.}
{\it A.J.} {\bf 110}, 2597. 

\bibitem[Tytler \& Davis 1993]{Tytler93}
 {\sc Tytler, D.\ \& Davis, C.} 1993
{On the Lack of Emission or Absorption in the Lyman Continua of Qsos:
 what is the Source of the Lyman Continuum Radiation, and where are the
 Broad Line Clouds?}
 {\it American Astronomical Society Meeting} {\bf 183}, 9106.

\bibitem[Ulvestad, Antonucci \& Goodrich (1995)]{Ulvestad95} 
{\sc Ulvestad, J.\ S., Antonucci, R.\ R.\ J.\ \& Goodrich, R.\ W.} 1995
{Radio properties of narrow-lined Seyfert 1 galaxies.}
{\it A.J.} {\bf 109}, 81.

\bibitem[van Bemmel \& Bertoldi 2001]{vanBemmel01}
{\sc van Bemmel, I. and Bertoldi, F.} preprint 2001
{Millimeter observations of radio-loud active galaxies.}
{\it astro-ph  0101137}

\bibitem[Vermeulen {\it et al.}\/ 1995]{Vermeulen95}
{\sc Vermeulen, R.\ C., Ogle, P.\ M., Tran, H.\ D., Browne, I.\ W.\ A.,
Cohen, M.\ H., Readhead, A.\ C.\ S., Taylor, G.\ B.\ \& Goodrich, R.\ W.} 1995
{When Is BL Lac Not a BL Lac?}
{\it ApJ. Lett.} {\bf 452}, L5. 

\bibitem[Vernet {\it et al.}\/ 2001]{Vernet00}
{\sc Vernet, J., Fosbury, R.\ A.\ E., Villar-Martin, M., Cohen, M.\ H.,
Cimatti, A., di Serego Alighieri, S., and Goodrich, R.\ W.} 2001
{\it Astron. Astrophys.} {\bf 366}, 7.

\bibitem[Vignati {\it et al.}\/ (1999)]{Vignati99}
{\sc Vignati, P.} {\it et al.}\/ 1999
{BeppoSAX unveils the nuclear component in NGC 6240.}
{\it AAP} {\bf 349}, L57. 

\bibitem[Walter \& Fink 1993]{Walter93}
 {\sc Walter, R.\ \& Fink, H.\ H.} 1993
{The Ultraviolet to Soft X-Ray Bump of SEYFERT-1 Type Active
Galactic Nuclei.}
 {\it AAP} {\bf 274}, 105.

\bibitem[Weaver 2000]{Weaver00}
  {\sc Weaver, K.\ A.} 2000
{Probing Dense Matter in the cores of AGN: Observations with RXTE and ASCA}
 To appear in ``Proceedings of X-ray Astronomy '99 -
Stellar Endpoints, AGN and the Diffuse Background," 2000.
G.\ Malaguti, G.\ Palumbo \& N.\ White (eds), Gordon \& Breach (Singapore)

\bibitem[Weymann, Morris, Foltz, \& Hewett(1991)]{1991ApJ...373...23W} 
{\sc Weymann, R.\ J., Morris, S.\ L., Foltz, C.\ B., \& Hewett, P.\ C.} 1991, 
{\it ApJ} {\bf 373}, 23.

\bibitem[Wills et al.(1992)]{1992ApJ...400...96W}
{\sc Wills, B.\ J., Wills, D., 
Evans, N.\ J., Natta, A., Thompson, K.\ L., Breger, M., \& Sitko, M.\ L.} 
1992, {\it ApJ} {\bf 400}, 96.

\bibitem[Wills and Brotherton (1995)]{Wills95}
{\sc Wills, B. J. and Brotherton, M. S.} 1995
{An Improved Measure of Quasar Orientation}
{\it ApJ} {\bf 448} L81.

\bibitem[Wolfe (1978)]{Wolfe78}
   {\sc Wolfe, A.\ M.} 1978
   {\it Pittsburgh Conference on BL Lac Objects}, Pittsburgh, Pa.,
   University of Pittsburgh, 1978.\ 439 p.
%\ (For individual items see A79-30027 to A79-30055).

\bibitem[Wrobel \& Lind 1990]{Wrobel90}
   {\sc Wrobel, J.\ M.\ \& Lind, K.\ R.} 1990
   {The double-lobed blazar 3C 371.}
   {\it ApJ.} {\bf 348}, 135.

\bibitem[Young {\it et al.\/} (1996)]{Young96}
   {\sc Young, S., Hough, J.\ H., Efstathiou, A., Wills, B.\ J.,
        Axon, D.\ J., Bailey, J.\ A.\ \& Ward, M.\ J.} 1996
   {Scattered broad optical lines in the polarized flux spectrum of the FR II
    galaxy 3C 321.}
   {\it MNRAS} {\bf 279}, L72.

\bibitem[Zheng {\it et al.}\/ (1998)]{Zheng98}
{\sc Zheng, W., Kriss, G.\ A., 
Telfer, R.\ C., Grimes, J.\ P.\ \& Davidsen, A.\ F.} 1998
{A Composite HST Spectrum of Quasars.}
{\it ApJ.} {\bf 492}, 855.

\end{thebibliography}
\end{document}